\documentclass[twocolumn,superscriptaddress,amsmath,amssymb,aps,longbibliography,prx,floatfix]{revtex4-2}
\usepackage[colorlinks=true,urlcolor=blue,citecolor=blue,linkcolor=blue]{hyperref}
\usepackage{graphicx}
\usepackage{dsfont}
\usepackage{booktabs}
\usepackage{amsmath}

\usepackage{braket}
\usepackage{dcolumn} % Align table columns on decimal point
\usepackage{makecell}

\usepackage{xcolor}   % for colored text
\usepackage{pifont}   % for \ding symbols
\usepackage{xspace}
\usepackage{multirow}

\newcommand{\cmark}{\textcolor{green!60!black}{\ding{51}}} % ✓
\newcommand{\xmark}{\textcolor{red!70!black}{\ding{55}}}   % ✗
\newcommand{\Eq}[1]{Eq.~(\ref{#1})}

\newcommand{\csp}{\texttt{CrystalFormer-CSP}\xspace}

\begin{document}

%\title{\texttt{CrystalFormer-CSP}: Generative Pretrained Transformer for Crystal Structure Prediction}
\title{\csp: Thinking Fast and Slow for Crystal Structure Prediction}

\author{Zhendong Cao}
\affiliation{Institute of Physics, Chinese Academy of Sciences, Beijing, China}
\affiliation{School of Physics, University of Chinese Academy of Sciences, Beijing, China}

\author{Shigang Ou}
\affiliation{Institute of Physics, Chinese Academy of Sciences, Beijing, China}
\affiliation{School of Physics, University of Chinese Academy of Sciences, Beijing, China}

\author{Lei Wang}
\email{wanglei@iphy.ac.cn}
\affiliation{Institute of Physics, Chinese Academy of Sciences, Beijing, China}

\begin{abstract}
Crystal structure prediction is a fundamental problem in materials science. We present \csp, an efficient framework that unifies data-driven heuristic and physics-driven optimization approaches to predict stable crystal structures for given chemical compositions. The approach combines pretrained generative models for space-group-informed structure generation and a universal machine learning force field for energy minimization. Reinforcement fine-tuning can be employed to further boost the accuracy of the framework. We demonstrate the effectiveness of \csp on benchmark problems and showcase its usage via web interface and language model integration.
\\
\begin{center}
    \vspace{0.25em}
    \setlength{\fboxsep}{5pt}
    \setlength{\fboxrule}{1pt}
    \large
    \fbox{%
      \parbox{0.6\linewidth}{%
        \centering
        Code: \url{https://github.com/deepmodeling/crystalformer-csp}\\[0.3em]
        Demo: \url{https://tinyurl.com/2ptdkz8w}%
      }%
    }
  \vspace{0.25em}
\end{center}
\end{abstract}

\maketitle 

\section{Introduction}

Crystal structure prediction (CSP) is a fundamental problem in materials science with broad applications spanning from superconductors to battery materials~\cite{woodley2019crystal, oganov2019structure}. The goal of CSP is to predict the spatial arrangement of atoms given the chemical formula, which is notoriously difficult due to the complex interplay between chemical bonding, geometric constraints, and thermodynamic stability.

Heuristics have historically played a crucial role in predicting and rationalizing crystal structures, especially before the advent of modern computational methods. Notable among these are the Pauling rules~\cite{Pauling1929}, which are a set of empirical guidelines derived from observations on a wide range of known inorganic crystals. The Pauling rules encapsulate chemical and geometric principles—such as coordination preferences, ionic size ratios, and charge neutrality—that help explain and anticipate how atoms pack and bond in stable arrangements. Other heuristics include the use of the Goldschmidt tolerance factor for perovskite formation and simple ionic radii considerations to estimate potential lattice configurations~\cite{Goldschmidt1926, sciadv.aav0693}. Another useful heuristic for crystal structure prediction is based on element substitution. This approach leverages the observation that many crystal structures are preserved when substituting chemically similar elements into known compounds. By identifying and transferring structural prototypes from well-characterized materials to new compositions with analogous chemical environments, one can generate reasonable initial guesses for unknown crystals. While these heuristic approaches are not universally predictive~\cite{George2020_PaulingRules_Limitations}, they provide valuable chemical intuition, narrow down plausible structural prototypes, and often serve as constraints or initial guesses in more systematic approaches to CSP.

Traditional computational approaches to CSP focus on identifying the most stable atomic arrangements by searching for low-energy configurations. These methods rely on a combination of local and global optimization algorithms to efficiently explore the potential energy landscape, enabling the prediction of crystal structures across a broad range of materials~\cite{wang2012calypso, pickard2011airss, glass2006uspex}. Recent advances in machine learning methods have greatly enhanced the capabilities of computational CSP. Machine learning force fields (MLFFs) model the potential energy surface of materials and provide a rapid way to relax crystal structures to local minima. This enables structure searches that are much more efficient than first-principles calculations~\cite{wang2023data}. Advances in universal MLFFs that balance accuracy and coverage across the periodic table~\cite{orb-v3, zhang2023dpa2, batatia2024foundation, zhang2025graph, yang2024mattersim, kim2024data} have enabled thorough exploration of chemical space~\cite{merchant2023scaling, schmidt2024improving, wang2025discovery}. 

However, the question of how to generate diverse yet reasonable initial crystal structures remains challenging~\cite{li2025selfoptimizingmachinelearningpotential}. As is well known in optimization, the quality of the initial guess can strongly affect the performance of subsequent optimization. Materials generative models offer a systematic solution to this problem, greatly extending conventional heuristics. These generative models capture heuristics as chemical patterns and generate novel structures by learning from datasets of known materials~\cite{cdvae, diffcsp, mattergen, flowmm, crystalformer, crystalflow, equicsp}. 

Directly applying these models to CSP tasks faces challenges, especially for large crystals with complex structures. For example, prediction accuracy drops rapidly from 78\% for crystals with at most 20 atoms per unit cell to 34\% for systems with up to 52 atoms per unit cell~\cite{diffcsp}. The CSP task is particularly demanding for systems with many atoms in the unit cell, as this increases both the search space and the computational cost of each optimization step. One way to scale up materials generative models to large complex systems is to fully exploit the natural inductive bias: the space group symmetries inherent in crystalline materials. Many seemingly complex crystals actually have very concise representations once symmetry constraints are taken into account.
%While compared to \cite{diffcsp}, the prediction centers around the space group symmetries, which is known to be a crucial inductive bias in CSP tasks~\cite{calypso, uspex, airss}. 
The symmetry inductive bias is exploited in \texttt{CrystalFormer}~\cite{crystalformer}, an autoregressive transformer for crystalline materials design and discovery. \texttt{CrystalFormer} models the conditional probability $p(x|g)$ to generate novel crystals $x$ in the given space group $g$. Therefore, it is designed for exploring a broad class of chemical composition and stoichiometry in the material space. 

The CSP task, however, requires generating a structure $x$ with a specific chemical formula $f$, i.e., sampling from the conditional probability distribution $p(x|f)$. To address this, we introduce \texttt{CrystalFormer-CSP}, a pretrained generative model for crystal structure prediction combined with universal MLFFs. The approach leverages an autoregressive transformer trained on stable crystal structures to generate initial candidate structures, followed by energy minimization with MLFFs and ranking according to energy above hull. Benchmark results demonstrate competitive performance with state-of-the-art CSP methods. Furthermore, the entire workflow can be enhanced through reinforcement learning (RL) to further boost performance.

The remainder of this paper is organized as follows: Section~\ref{sec:method} describes the \csp methodology with a comparison to related works in CSP; Section~\ref{sec:usage} introduces the usage of \csp via either web interface or language model integration; Section~\ref{sec:results} presents benchmark results on CSP tasks; Section~\ref{sec:discussion} concludes with a discussion of implications and future directions.

\section{Methodology}
\label{sec:method} 

\begin{figure*}[t]
    \centering
    \includegraphics[width=\textwidth]{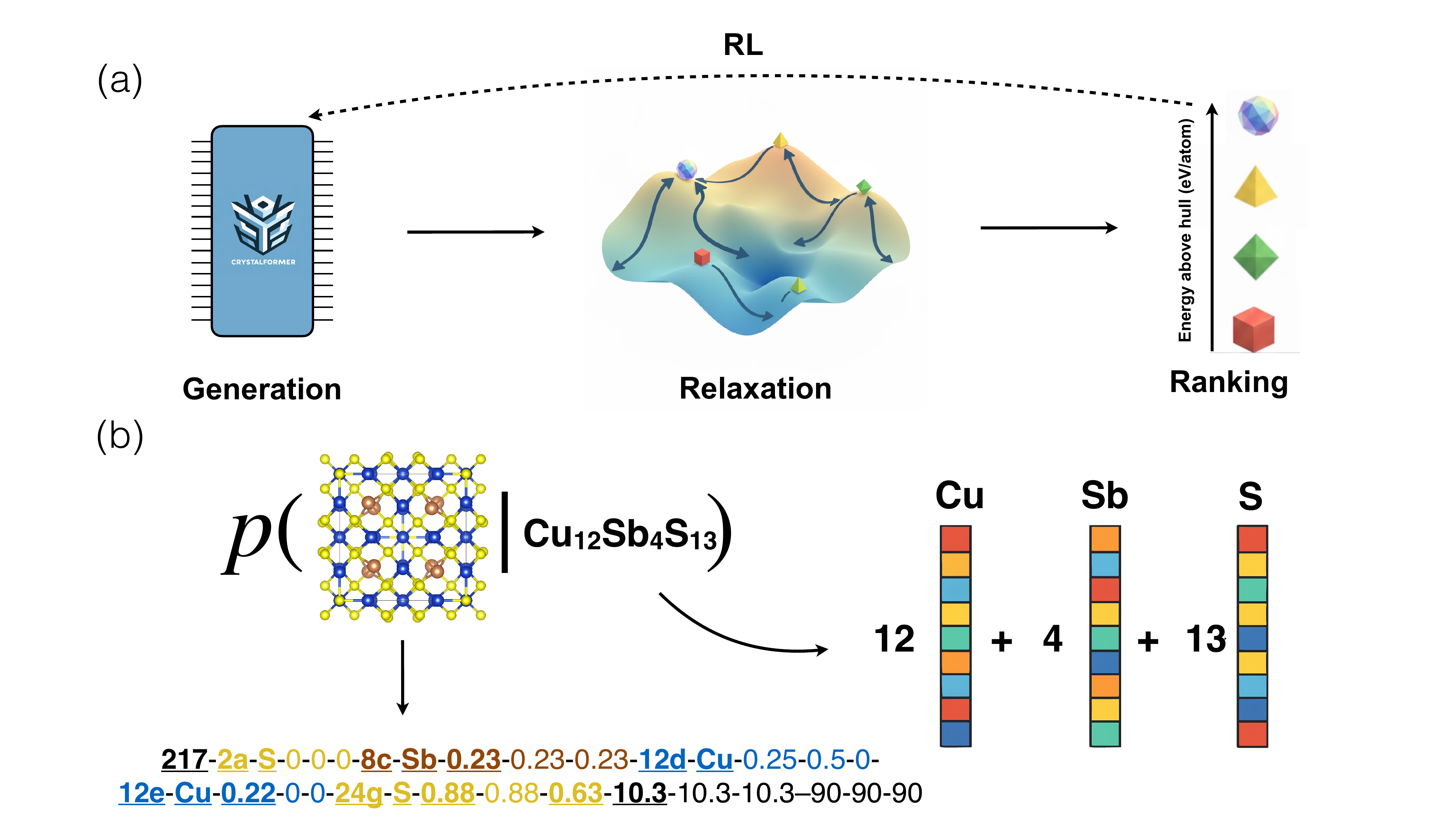}
    \caption{
        (a) The \csp framework. The process starts from a user-specified chemical formula and proceeds as follows: \ding{192} \textbf{Generation}: the pretrained autoregressive transformer samples a batch of crystal structure candidates; \ding{193} \textbf{Relaxation}: the machine learning force field relaxes the structures and provides energy estimates of the final structures; \ding{194} \textbf{Ranking}: the relaxed candidates are ranked according to their energy above hull ($E_{\mathrm{hull}}$), providing an assessment of their thermodynamic stability. Optionally, one can further fine-tune the generative model with reinforcement learning using the relaxed energy as the reward signal. 
        (b) The generation step samples from the conditional probability distribution of crystal structure given a chemical formula. The chemical formula is represented as a weighted summation of learned embedding vectors for the chemical elements. The crystal structure is represented as a sequence that consists of space group number, Wyckoff letters, atom species, fractional coordinates, and lattice parameters. In practice, only these underlined bold tokens need to be sampled; the remaining ones are fixed by the preconditions.
         }
    \label{fig:concept}
\end{figure*}

\csp employs a three-stage approach for CSP, as shown in Fig.~\ref{fig:concept}(a). First, a pretrained autoregressive transformer samples initial crystal structure candidates based on the given chemical formula. Second, an MLFF relaxes these structures to lower energy states via structure relaxation. Finally, the relaxed candidates are ranked according to their energy above hull ($E_{\mathrm{hull}}$), providing an assessment of their thermodynamic stability.

The core of this workflow is a generative model that models the conditional probability of a crystal structure given a chemical formula, similar to how language models sample answers given prompts. \texttt{CrystalFormer-CSP} is an autoregressive transformer designed for generating inorganic crystal structures with space group symmetries. The model takes a chemical formula as input and outputs a crystal structure. As shown in Fig.~\ref{fig:concept}(b), the chemical formula is represented as a weighted sum of atomic embedding vectors. The crystal structure is represented as a sequence of space group, Wyckoff positions, atom species, fractional coordinates, and lattice parameters. Note that \texttt{CrystalFormer} imposes alphabetical ordering on the Wyckoff letters in the sequence~\cite{crystalformer}, consistent with sampling the positions of symmetry-inequivalent atoms from higher to lower symmetry sites. Additionally, fractional coordinates and lattice parameters are modeled as continuous variables following mixtures of periodic von Mises and Gaussian distributions, respectively. These variables are constrained to be consistent with the given space group and Wyckoff letters.
%color{red}{In addition, the Wyckoff letters are constrained to be in the alphabet of the given space group. And they are appeared in the alphabet order, indicating that one samples the position of symmetry inequivalent atoms from higher to lower symmetry sites. The fractional coordinates and lattice parameters are also constrained to be consistent with the given space group and Wyckoff letters.}

%architecture
Compared to the original \texttt{CrystalFormer} architecture~\cite{crystalformer}, the \csp model places the chemical formula as the prefix of the entire sequence. The formula representation is given by a weighted sum of atomic embedding vectors, making it invariant to the order of atoms in the formula. The remaining part of the sequence follows the same order as the original \texttt{CrystalFormer} design. Generation is constrained so that atoms are drawn from the given formula. This enables the generative model to efficiently generate complex crystal structures with the specified chemical formula and space-group symmetry.
%training
We train the model using maximum likelihood estimation on a filtered version of the Alexandria dataset~\cite{alexandria, cavignac2025aidriven}. The dataset contains approximately 1.7 million inorganic crystalline materials with energy above the convex hull less than 0.1 eV/atom and no more than 20 Wyckoff sites in the conventional cell. The dataset covers 89 chemical elements. We use an 80-10-10 split for training, validation, and test datasets, and release the dataset at \href{https://huggingface.co/datasets/zdcao/alex-20s}{https://huggingface.co/datasets/zdcao/alex-20s}.

\begin{figure}[h]
    \centering
    \includegraphics[width=\columnwidth]{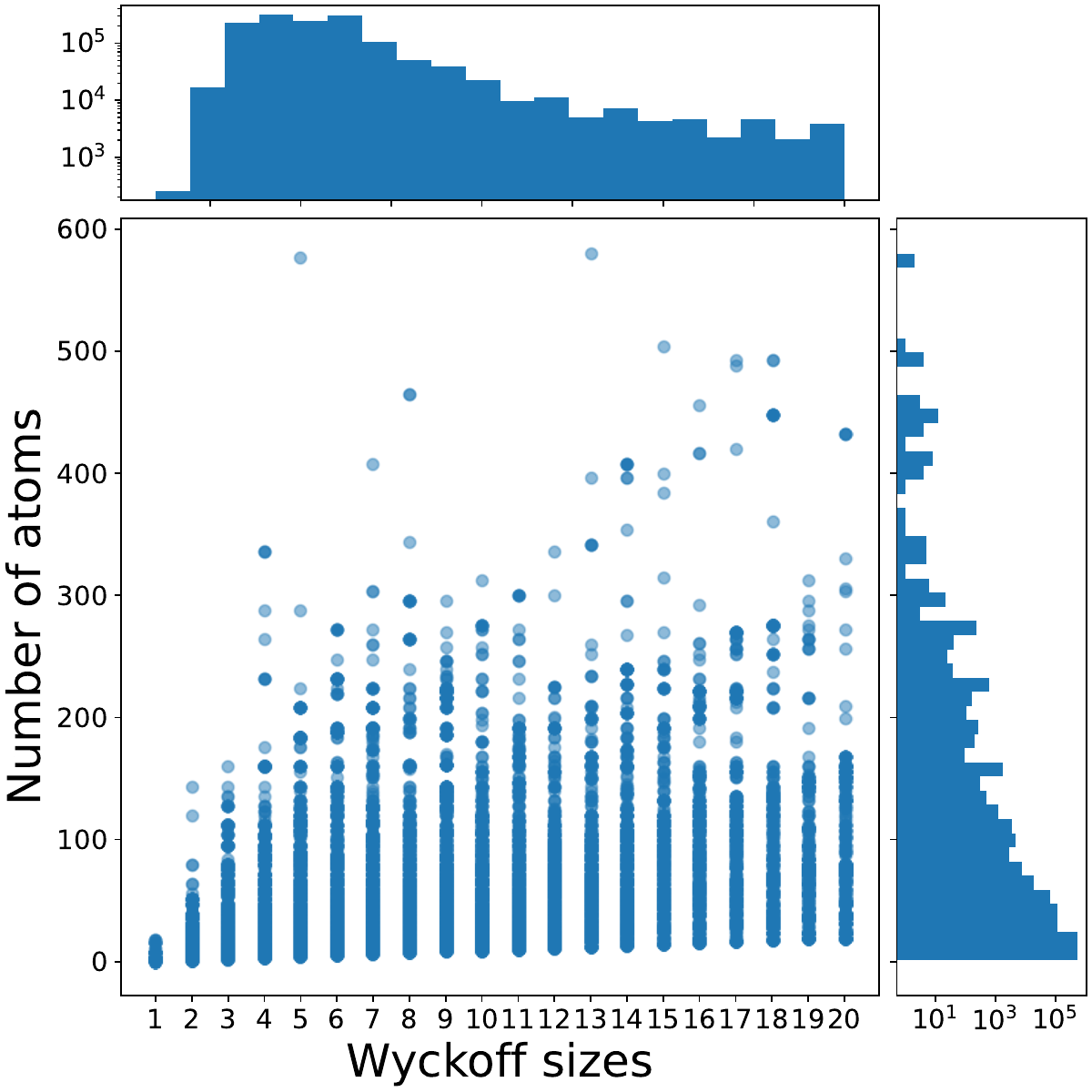}
    \caption{
    The correlation plot of the number of atoms in the conventional cell and the number of Wyckoff sites. The histograms are shown in the side panels. 
    }
    \label{fig:atomcount_hist}
\end{figure}

Figure~\ref{fig:atomcount_hist} shows the correlation between the number of atoms in the conventional cell and Wyckoff size across the training dataset. The Wyckoff size is defined as the number of Wyckoff sites occupied by symmetry-inequivalent atoms. For example, the Wyckoff size of the Cu$_{12}$Sb$_4$S$_{13}$ structure shown in Fig.~\ref{fig:concept}(b) equals to 5, with five occupied Wyckoff positions: 2a, 8c, 12d, 12e, and 24g. This number is much smaller than 58: the total number of atoms in the unit cell, which equals to the sum of multiplicities of the occupied Wyckoff positions. Figure~\ref{fig:atomcount_hist} shows that the Wyckoff sequences are much shorter than the number of atoms over the whole dataset, demonstrating its efficiency in representing complex crystal structures. Similar observation has also been made for Pearson's Crystal Data (the Crystal Structure Database for Inorganic Compounds)~\cite{hornfeck2022combinatorics}. 

The raw training data in CSV format is 4.6 GB. After extracting essential crystal structure data (space group, Wyckoff sequence, atom species, fractional coordinates of each symmetry-inequivalent atom, and lattice parameters), the size is reduced to 593 MB. The final model checkpoint is 52 MB, achieving approximately a tenfold compression ratio~\cite{Deletang2024_LMIsCompression}. At inference time, the generative model may recall structures from the training data as predictions. For unseen formulas, the model samples crystals following the same distribution as low-energy training samples. 

%inference
At inference time, we use the pretrained model to sample a batch of crystal structures conditioned on the given chemical formula. We then filter candidate crystals to match the correct chemical stoichiometry. Next, we relax the structures using an MLFF~\cite{riebesell2025framework}. The relaxation process modifies atomic positions and lattice parameters, which may alter the space group. By combining a space-group-informed crystal generative model for symmetric structure generation with a universal force field for structure relaxation, the framework handles both discrete and continuous degrees of freedom in CSP in a unified manner. 

Finally, we rank the relaxed structures based on their energy above hull ($E_\textrm{hull}$) relative to the convex hull~\cite{alexandria, cavignac2025aidriven}. For a given composition, the convex hull is constructed from all known stable structures formed with the specified chemical species, and $E_\textrm{hull}$ is the energy difference between the predicted structure and the hull. Since we filter by formula before relaxation, all relaxed structures have the same stoichiometry. Therefore, ranking by $E_\textrm{hull}$ is equivalent to ranking by energy per formula. However, the benefit of computing $E_\textrm{hull}$ is that its value more directly assesses the thermodynamic stability of the structures. Structures with $E_\textrm{hull} < 0$ are considered thermodynamically stable, while those with small positive values (typically $< 0.1$ eV/atom) may be metastable and experimentally accessible~\cite{sun2016thermodynamic}. 

Building on this pipeline, we can employ RL to further fine-tune the generative model for specific chemical formulas to boost CSP accuracy. As noted in Ref.~\cite{silver2025welcome}, once available training data is saturated, further improvements require agents to learn directly through interaction with the environment. The RL procedure corresponds to a feedback loop shown in Figure~\ref{fig:concept}(a). As a bottom line, running the pipeline repeatedly already allows one to find stable crystals through multiple rounds of sampling. RL guides the search toward low-energy structures by shifting the distribution of generated samples with a feedback loop to the generative model.  

We use the estimated $E_{\textrm{hull}}$ as the reward and fine-tune the generative model to \emph{maximize} the following objective function~\cite{crystalformer-rl}: 
\begin{equation}
  \label{eq:rl}
  \mathcal{L}_{\theta} = \mathop{\mathbb{E}}_{x \sim p_{\theta}(x|f)} \left[-E_{\textrm{hull}}(x^\prime) \right] + \mathop{\mathbb{E}}_{x \sim \textrm{buffer}} \left[ \ln p_{\theta}(x|f) \right], 
\end{equation}
where $p_{\theta}(x|f)$ is the model likelihood for generating structure $x$ given chemical formula $f$. Energy relaxation modifies the sampled structures from $x$ to $x^\prime$, which are then used to evaluate energy above hull. The second term corresponds to maximum likelihood estimation on an experience buffer collected during the RL process. The buffer contains samples that lead to the lowest energy above hull after relaxation. Note that the buffer stores generated samples rather than relaxed samples, aiming to smoothly shift the pretrained distribution without catastrophic forgetting of pretrained knowledge~\cite{shenfeld2025rlsrazoronlinereinforcement}. RL tunes the neural network parameters $\theta$ of the generative model to maximize the objective function in \Eq{eq:rl}.

The two terms in the objective function focus on the mean and extreme of the energy above hull, respectively. Viewed as an optimization method, the RL approach relates to recent advances in high-dimensional optimization using tensor trains as probabilistic models~\cite{PROTES, sozykin2025highdimensionaloptimizationlowrank}. The key difference is that the pretrained autoregressive transformer provides an excellent initial policy for CSP tasks. More importantly, we conjecture that pretraining reshapes the optimization landscape over neural network parameters to be more friendly compared the original potential energy surface of the crystals. For example, a single neural network parameter in the crystal generative model may control a number of atom species and coordinates. Policy gradient updates of neural network parameters according to the reward signal then mutates the generated crystal samples in a nonlocal manner. 

The RL loss function mirrors that used for fine-tuning large language models~\cite{Ouyang2022InstructGPT} and can be optimized using the same proximal policy optimization algorithm~\cite{schulman2017ppo}. Similar to advances in enhancing reasoning ability of large language models, we expect reinforcement fine-tuning to be useful for challenging CSP instances. Methodologically, applying policy gradient-based RL to CSP provides an alternative to evolutionary structure search. The space-group-informed autoregressive model provides a unified way to parametrize the search space with both discrete (space group, Wyckoff letters) and continuous (fractional coordinates, lattice parameters) degrees of freedom. 

Eventually, first-principles calculations or experiments can verify the discovered structures. These verification results can also serve as RL rewards to fine-tune the generation process and further enhance its accuracy. 
%On the other hand, it is also possible to feed the generated structures to existing CSP methods~\cite{wang2012calypso,oganov2011uspex,pickard2011airss}.

\section{Usage}
\label{sec:usage}

We discuss practical usage via web interface and language model integration. These provide convenient interfaces for using \csp. Advanced usage, which allows direct work with the code and model, is available at \href{https://github.com/deepmodeling/crystalformer-csp}{github.com/deepmodeling/crystalformer-csp} with full control.

\begin{figure*}[t]
  \centering
    \includegraphics[width=1\textwidth]{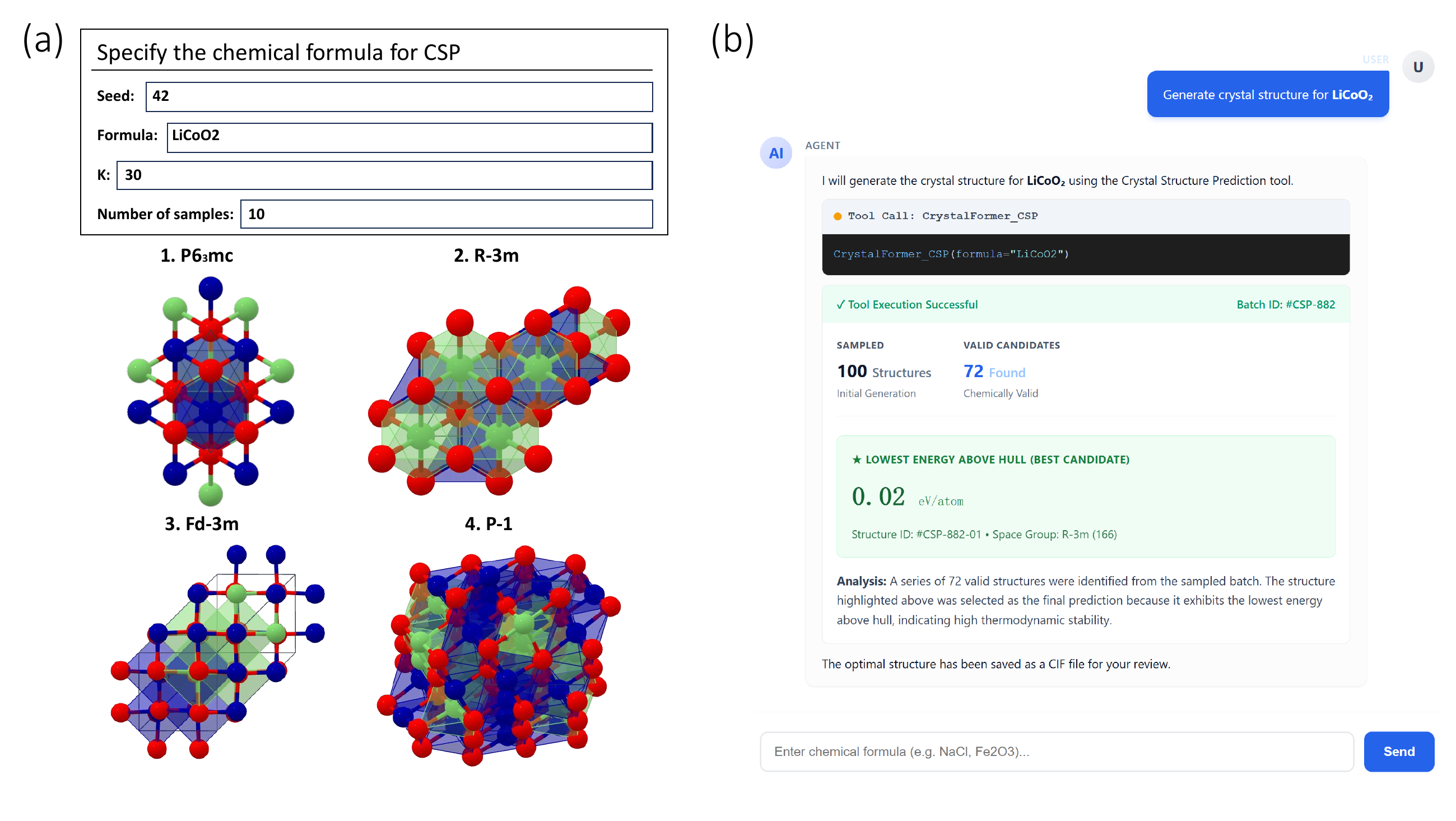}
  \caption{Schematic illustrations of (a) Colab notebook interface, (b) Natural language model integration.
  }
  \label{fig:usage}
\end{figure*}

\subsection{Colab notebook}

We provide the workflow shown in Fig.~\ref{fig:concept} as a Colab notebook. Users can input a chemical formula into a text box, click run, and receive 5 structures with estimates of their energy above hull. Inspired by ColabFold~\cite{Mirdita2022}, our goal is to develop \csp into ColabCSP---an easy-to-use, free, and accessible platform for CSP.

\subsection{Language model integration}

We have integrated \csp into a language model agent via the Model Context Protocol (MCP). This enables users to perform CSP tasks through intuitive natural language prompts, streamlining structure generation, relaxation, and analysis in a seamless conversational workflow.

% We present detailed case studies for several challenging systems to demonstrate the capabilities of \texttt{CrystalFormer-CSP}:

\section{Results}
\label{sec:results}

We first evaluate the performance of \csp on a wide range of CSP tasks, then analyze its application through case studies.

\subsection{Space group prediction accuracy}

The space group is the first token predicted by the model given the chemical formula. On the 230-class space-group classification task, our model achieves an average negative log-likelihood (NLL) of 1.78 nats on the validation dataset, corresponding to a perplexity of 5.92. In other words, the predictive distribution is as concentrated as a uniform distribution over approximately six effective classes rather than all 230 (uniform baseline: NLL = $\ln 230 \approx 5.44$, perplexity = 230). 

This translates to robust top-1 performance (substantially above random) and strong practical discriminability among space groups. Together, these metrics indicate that the network has learned salient crystallographic symmetry cues. Note that atom embeddings also appear in later stages of crystal structure sampling. As shown in Ref.~\cite{crystalformer}, learned atom embedding vectors align for chemically similar elements. We believe this provides a strong regularization effect on the space group predictor.

The prediction accuracy for space groups on the test dataset is shown in Table~\ref{tab:spacegroup}. Achieving excellent performance in space group prediction provides a valuable starting point for crystal structure prediction. 

\begin{table}[h]
  \centering
  \begin{tabular}{c c}
  \hline
  \textbf{Top-K} & \textbf{Accuracy (\%)} \\
  \hline
  1  & 52.7 \\
  10 & 90.7 \\
  20 & 96.2 \\
  30 & 98.1 \\
  40 & 98.9 \\
  50 & 99.3 \\
  \hline
  \end{tabular}
  \caption{Top-K accuracy in predicting the space group of materials in the test dataset.}
  \label{tab:spacegroup}
\end{table}

In practical sampling, to encourage diversity of space groups, we uniformly sample from the top $K$ space groups rather than directly following predicted space group logits. Table~\ref{tab:spacegroup} shows that with $K=40$, we achieve 98.9\% accuracy in space group prediction.

\subsection{Crystal structure prediction benchmarks}

We evaluate \csp on two benchmark datasets used in CSP~\cite{shotguncsp}, spanning diverse chemical spaces in terms of space groups, chemical elements, and number of atoms. We have verified that all ground truth structures are stable with respect to the \texttt{orb-v3-conservative-inf-mpa} force field~\cite{orb-v3}. Nevertheless, the MLFF may assign these stable structures energy above hull values that deviate from zero. 

For each structure, we run sampling until collecting 500 structures with matching formula. We then relax the structures using the \texttt{orb-v3-conservative-inf-mpa} force field in batch mode~\cite{BatchRelaxer}. The relaxation protocol uses a force convergence criterion of 0.01 eV/\AA. Finally, we compare the ground truth structure with relaxed structures having $E_\textrm{hull} < 1.0$ eV/atom to determine prediction success. Although precise stoichiometric ratios are not strictly enforced during generation, we find that the model generates a reasonable fraction of structures with the desired stoichiometry. Consequently, the sampling process takes a small fraction of the time compared to relaxation and energy evaluation.

The performance of \csp reported in Tables~\ref{tab:cspbench-datasetI} and~\ref{tab:cspbench-datasetII} shows it is competitive with state-of-the-art CSP methods. It provides accurate predictions regardless of whether the structure is in the training dataset. In cases where predicted structures do not match the ground truth, we find that relaxed energies can even be lower than the ground truth. Several structures in dataset II have Wyckoff sizes exceeding \csp's context window of 20, leading to failure. Overall, match rates on datasets I and II (82.5\% and 80\%) exceed the performance of 77.5\% and 78\% achieved in~\cite{shotguncsp}. \csp differs from Ref.~\cite{shotguncsp} in being an \emph{end-to-end} model that predicts space group, Wyckoff sequence, atom species, fractional coordinates, and lattice parameters in one unified model, rather than dividing them into individual components. This enables joint optimization of the entire prediction pipeline, as we demonstrate in the following section. 

We also note that \csp successfully recovers all 15 structures selected by DiffCSP from their test dataset~\cite{diffcsp}, demonstrating the effectiveness of our pretrained approach that focuses on finding stable structures regardless of train/test splits.

%\lw{hows is diffcsp 's performance in csp ???}
%\zdcao{For the MP-20 test dataset, the DiffCSP achieves a top-1 accuracy of 51.49\% and a top-20 accuracy of 77.93\%.}

\begin{table*}[htbp]
\centering
\caption{Benchmark of \texttt{CrystalFormer-CSP} performance for representative materials collected in Ref.~\cite{shotguncsp} (dataset I). A match means there is a match to the ground truth structure with the default setup of \texttt{StructureMatcher} in PyMatGen. For those cases where \csp does match the given structure, we list the energy above hull of the reference structure and the lowest value of discovered structures evaluated with the \texttt{orb-v3-conservative-inf-mpa} force field. 
}
\resizebox{\textwidth}{!}{
\begin{tabular}{lrcccccccc}
  \toprule
\multirow{2}{*}{\textbf{Composition}}
& \multirow{2}{*}{\textbf{Space group}}
& \multirow{2}{*}{\textbf{Number of atoms}}
& \multirow{2}{*}{\textbf{Wyckoff size}}
& \multicolumn{2}{c}{\textbf{Match}}
& \multirow{2}{*}{\textbf{Training}}
& \multicolumn{3}{c}{\textbf{$E_\mathrm{hull}$ (eV/atom) $\downarrow$}}
\\ 
\cmidrule(lr){5-6} \cmidrule(lr){8-10}
& & & & \textbf{w/o RL} & \textbf{w/ RL} & & \textbf{Reference} & \multicolumn{2}{c}{\textbf{Minimum}} \\
\cmidrule(lr){9-10}
& & & & & & & & \textbf{w/o RL} & \textbf{w/ RL} \\
\midrule
Si                   & $Fd\bar{3}m$ & 2   & 1  & \cmark & & \cmark & &  &  \\
GaAs                 & $F\bar{4}3m$ & 2   & 2  & \cmark & & \cmark & &  &  \\
ZnO                  & $P6_3mc$     & 4   & 2  & \cmark & & \cmark &  &  &  \\
BN                   & $P6_3/mmc$   & 4   & 2  & \cmark & & \cmark & &  &  \\
C                    & $R\bar{3}m$  & 4   & 2  & \cmark & & \cmark & & & \\
Ba(FeAs)$_2$         & $I4/mmm$     & 5   & 3  & \cmark & & \cmark & & & \\
Bi$_2$Te$_3$         & $R\bar{3}m$  & 5   & 3  & \cmark & & \cmark & & & \\
VO$_2$               & $P4_2/mnm$   & 6   & 2  & \cmark & & \cmark & &  &  \\
SiO$_2$              & $I\bar{4}2d$ & 6   & 2  & \cmark & & \cmark & & & \\
La$_2$CuO$_4$        & $I4/mmm$     & 7   & 4  & \cmark & & \xmark & & & \\
LiPF$_6$             & $R\bar{3}$   & 8   & 3  & \cmark & & \cmark & & & \\
SrTiO$_3$            & $I4/mcm$     & 10  & 4  & \cmark & & \xmark & & & \\
CaCO$_3$             & $R\bar{3}c$  & 10  & 3  & \cmark & & \cmark & & & \\
Al$_2$O$_3$          & $R\bar{3}c$  & 10  & 2  & \cmark & & \cmark & & & \\
ZrO$_2$              & $P2_1/c$     & 12  & 3  & \cmark & & \xmark & & & \\
TiO$_2$              & $C2/m$       & 12  & 6  & \xmark & \cmark & \cmark & 0.0103 & \textbf{0.0097} & \textbf{0.0097}\\
ZrTe$_5$             & $Cmcm$       & 12  & 4  & \cmark & & \cmark & & & \\
V$_2$O$_5$           & $Pmmn$       & 14  & 4  & \cmark & & \cmark & & & \\
Si$_3$N$_4$          & $P6_3/m$     & 14  & 3  & \cmark & & \cmark & & & \\
Fe$_3$O$_4$          & $Fd\bar{3}m$ & 14  & 5  & \cmark & & \xmark & & & \\
Mn(FeO$_2$)$_2$      & $Fd\bar{3}m$ & 14  & 3  & \cmark & & \cmark & & & \\
ZnSb                 & $Pbca$       & 16  & 2  & \cmark & & \cmark & & & \\
CoSb$_3$             & $Im\bar{3}$  & 16  & 2  & \cmark & & \xmark & & & \\
LiCoO$_2$            & $R\bar{3}m$  & 16  & 3  & \cmark & & \cmark & & & \\
LiBF$_4$             & $P3_121$     & 18  & 4  & \cmark & & \cmark & & & \\
Y$_2$Co$_{17}$       & $R\bar{3}m$  & 19  & 5  & \cmark & & \cmark & & & \\
CsPbI$_3$            & $Pnma$       & 20  & 5  & \cmark & & \cmark & & & \\
GeH$_4$              & $P2_12_12_1$ & 20  & 5  & \xmark & \xmark & \xmark & 0.062 & 0.011 & \textbf{-0.015}\\
NaCaAlPHO$_5$F$_2$   & $P2_1/m$     & 24  & 10 & \cmark & & \cmark & & & \\
LiFePO$_4$           & $Pnma$       & 28  & 6  & \cmark & & \cmark & & & \\
Cu$_{12}$Sb$_4$S$_{13}$ & $I\bar{4}3m$ & 29 & 5 & \cmark & & \xmark & & & \\
Li$_3$PS$_4$         & $Pnma$       & 32  & 6  & \xmark & \cmark& \cmark & 0.099 & \textbf{0.082} & \textbf{0.082}\\
MgB$_7$              & $Imma$       & 32  & 7  & \xmark &\xmark & \cmark & \textbf{-0.003} & 0.193 & 0.020\\
Li$_4$Ti$_5$O$_{12}$ & $C2/c$       & 42  & 12 & \cmark & & \cmark & & & \\
Ba$_2$CaSi$_4$(BO$_7$)$_2$ & $I\bar{4}2m$ & 46 & 8 & \xmark &\cmark & \cmark & 0.0098 & 0.1097 & \textbf{0.0097}\\
Ag$_8$GeS$_6$        & $Pna2_1$     & 60  & 15 & \xmark & \cmark& \cmark & 0.068 & \textbf{0.065} & \textbf{0.065}\\
Nd$_2$Fe$_{14}$B     & $P4_2/mnm$   & 68  & 9  & \cmark & & \xmark & & & \\
Cd$_3$As$_2$         & $I4_1/acd$   & 80  & 6  & \xmark &\cmark & \xmark & 0.0031 & 0.0028 & \textbf{0.0025}\\
Y$_3$Al$_5$O$_{12}$  & $Ia\bar{3}d$ & 80  & 4  & \cmark & & \cmark & & & \\
Ca$_{14}$MnSb$_{11}$ & $I4_1/acd$   & 104 & 9  & \cmark & & \cmark & & & \\
\midrule
\textbf{Overall} &  & & &  33/40=82.5\% & 38/40=95\% \\
\bottomrule
\label{tab:cspbench-datasetI}
\end{tabular}
}
\end{table*}

\begin{table*}[htbp]
\centering
\caption{Benchmark of \texttt{CrystalFormer-CSP} performance for representative materials collected in Ref.~\cite{shotguncsp} (dataset II).
}
\resizebox{\textwidth}{!}{
\begin{tabular}{lrcccccccc}
  \toprule
\multirow{2}{*}{\textbf{Composition}}
& \multirow{2}{*}{\textbf{Space group}}
& \multirow{2}{*}{\textbf{Number of atoms}}
& \multirow{2}{*}{\textbf{Wyckoff size}}
& \multicolumn{2}{c}{\textbf{Match}}
& \multirow{2}{*}{\textbf{Training}}
& \multicolumn{3}{c}{\textbf{$E_\mathrm{hull}$ (eV/atom) $\downarrow$ }}
\\ 
\cmidrule(lr){5-6} \cmidrule(lr){8-10}
& & & & \textbf{w/o RL} & \textbf{w/ RL} & & \textbf{Reference} & \multicolumn{2}{c}{\textbf{Minimum}} \\
\cmidrule(lr){9-10}
& & & & & & & & \textbf{w/o RL} & \textbf{w/ RL} \\
\midrule
CsCl                & $Fm\bar{3}m$ & 2  & 2  & \cmark & & \cmark & &  &  \\
MnAl                & $P4/mmm$     & 2  & 2  & \cmark & & \cmark & &  &  \\
HoHSe               & $P\bar{6}m2$ & 3  & 3  & \cmark & & \cmark & &  &  \\
ErCdRh$_2$          & $Fm\bar{3}m$ & 4  & 3  & \cmark & & \xmark & &  &  \\
Eu$_2$MgTl          & $Fm\bar{3}m$ & 4  & 3  & \cmark & & \cmark & &  &  \\
Pm$_2$NiIr          & $Fm\bar{3}m$ & 4  & 3  & \cmark & & \xmark & &  &  \\
VPt$_3$             & $I4/mmm$     & 4  & 3  & \cmark & & \cmark & &  &  \\
Gd(SiOs)$_2$        & $I4/mmm$     & 5  & 3  & \cmark & & \cmark & &  &  \\
LaAl$_3$Au          & $I4mm$       & 5  & 4  & \cmark & & \cmark & &  &  \\
U$_2$SbN$_2$        & $I4/mmm$     & 5  & 3  & \cmark & & \xmark & &  &  \\
MnGa(CuSe$_2$)$_2$  & $I\bar{4}$   & 8  & 5  & \cmark & & \cmark & &  &  \\
SmZnPd              & $P\bar{6}2m$ & 9  & 4  & \cmark & & \cmark & &  &  \\
Sn(TePd$_3$)$_2$    & $I4mm$       & 9  & 7  & \xmark & \cmark & \cmark & 0.015 & 0.041 & \textbf{0.013}\\
V$_5$S$_4$          & $I4/m$       & 9  & 3  & \cmark & & \cmark & &  &  \\
Cs$_3$InF$_6$       & $Fm\bar{3}m$ & 10 & 4  & \cmark & & \xmark & &  &  \\
Eu(CuSb)$_2$        & $P4/nmm$     & 10 & 5  & \cmark & & \cmark & &  &  \\
Rb$_2$TlAgCl$_6$    & $Fm\bar{3}m$ & 10 & 4  & \cmark & & \cmark & &  &  \\
Ca$_3$Ni$_7$B$_2$   & $R\bar{3}m$  & 12 & 5  & \cmark & & \cmark & &  &  \\
DyPO$_4$            & $I4_1/amd$   & 12 & 3  & \cmark & & \cmark & &  &  \\
LaSiIr              & $P2_13$      & 12 & 3  & \cmark & & \cmark & &  &  \\
SmVO$_4$            & $I4_1/amd$   & 12 & 3  & \cmark & & \cmark & &  &  \\
VCl$_5$             & $P\bar{1}$   & 12 & 6  & \xmark & \xmark& \cmark & -0.487 & -0.518 & \textbf{-0.520}\\
YbP$_5$             & $P2_1/m$     & 12 & 4  & \cmark & & \xmark & &  &  \\
Eu(Al$_2$Cu)$_4$    & $I4/mmm$     & 13 & 4  & \cmark & & \cmark & &  &  \\
Zr$_4$O             & $R\bar{3}$   & 15 & 4  & \xmark & \xmark& \xmark & 0.013    & -0.002 & \textbf{-0.003} \\
Ba$_3$Ta$_2$NiO$_9$ & $P\bar{3}m1$ & 15 & 6  & \cmark & & \cmark & &  &  \\
K$_2$Ni$_3$S$_4$    & $Fddd$       & 18 & 4  & \cmark & & \cmark & &  &  \\
Sr(ClO$_3$)$_2$     & $Fdd2$       & 18 & 5  & \cmark & & \cmark & &  &  \\
LiSm$_2$IrO$_6$     & $P2_1/c$     & 20 & 6  & \cmark & & \cmark & &  &  \\
Pr$_2$ZnPtO$_6$     & $P2_1/c$     & 20 & 6  & \cmark & & \cmark & &  &  \\
Sc$_2$Mn$_{12}$P$_7$& $P\bar{6}$   & 21 & 9  & \cmark & & \xmark & &  &  \\
LaSi$_2$Ni$_9$      & $I4_1/amd$   & 24 & 4  & \cmark & & \xmark & &  &  \\
CeCu$_5$Sn          & $Pnma$       & 28 & 6  & \cmark & & \cmark & &  &  \\
LiP(HO$_2$)$_2$     & $Pna2_1$     & 32 & 8  & \xmark & \cmark & \xmark & 0.008 & 0.017 & \textbf{0.005}\\
Y$_4$Si$_5$Ir$_9$   & $P6_3/mmc$   & 36 & 7  & \cmark & & \xmark & &  &  \\
Mg$_3$Si$_2$H$_4$O$_9$ & $P6_3cm$ & 36 & 8  & \cmark & & \cmark & &  &  \\
Na(WO$_3$)$_9$      & $R\bar{3}$   & 37 & 9  & \cmark & & \cmark & &  &  \\
Sm$_6$Ni$_{20}$As$_{13}$ & $P\bar{6}$ & 39 & 15 & \cmark & & \xmark & &  &  \\
BaCaGaF$_7$         & $P2/c$       & 40 & 11 & \cmark & & \cmark & &  &  \\
Tm$_{11}$Sn$_{10}$  & $I4/mmm$     & 42 & 9  & \cmark & & \cmark & &  &  \\
AlH$_{12}$(ClO$_2$)$_3$ & $R\bar{3}c$  & 44 & 5  & \cmark & & \cmark & &  &  \\
K$_2$ZrSi$_2$O$_7$  & $P2_1/c$     & 48 & 12 & \cmark & & \cmark & &  &  \\
LiZr$_2$(PO$_4$)$_3$ & $P2_1/c$    & 72 & 18 & \cmark & & \cmark & &  &  \\
K$_5$Ag$_2$(AsSe$_3$)$_3$ & $Pnma$  & 76 & 13 & \xmark &\xmark & \xmark & \textbf{-0.220}    & -0.183 &   -0.200   \\
Be$_{17}$Ru$_3$     & $Im\bar{3}$  & 80 & 7  & \cmark & & \cmark & &  &  \\
Cu$_3$P$_8$(S$_2$Cl)$_3$ & $Pnma$  & 80 & 11 & \xmark &\xmark & \cmark & \textbf{-0.045}    & 0.067 & -0.001     \\
Al$_2$CoO$_4$       & $P3m1$       & 84 & 48 & \xmark &\xmark & \xmark & 0.056    & \textbf{0.039} &  \textbf{0.039}   \\
Li$_6$V$_3$P$_8$O$_{29}$ & $P1$    & 92 & 92 & \xmark &\xmark & \xmark & 0.040    & 0.030 &    \textbf{0.029}  \\
ReBi$_3$O$_8$       & $P2_13$      & 96 & 12 & \xmark &\cmark & \cmark & 0.012   & 0.056 &  \textbf{0.011}   \\
Na$_5$FeP$_2$(O$_4$F)$_2$ & $Pbca$ & 288 & 37 & \xmark &\xmark & \xmark & \textbf{-0.022}   & -0.020 &  -0.020   \\
\midrule
\textbf{Overall} & & & & 40/50=80\%  & 43/50=86\%\\
\bottomrule
\label{tab:cspbench-datasetII}
\end{tabular}
}
\end{table*}

\subsection{Reinforcement learning for crystal structure prediction}

To assess the impact of RL fine-tuning on CSP performance, we applied the RL procedure to all failed cases in the benchmark datasets. We use a batch size of 500 for samples drawn from \csp with matching formula and an experience buffer size of 50. As shown in Tables~\ref{tab:cspbench-datasetI} and~\ref{tab:cspbench-datasetII}, RL significantly improves match rates from 82.5\% to 95\% for dataset I and from 80\% to 86\% for dataset II. Notably, for all tested structures, RL either reduces or maintains the minimum $E_{\textrm{hull}}$ value, aligning with the fundamental goal of CSP: finding the lowest energy configurations. 

Interestingly, in many cases the minimal predicted $E_{\textrm{hull}}$ values obtained by RL are lower than those of reference structures: TiO$_2$, GeH$_4$, Li$_3$PS$_4$, Ba$_2$CaSi$_4$(BO$_7$)$_2$, Ag$_8$GeS$_6$, Cd$_3$As$_2$, Sn(TePd$_3$)$_2$, VCl$_5$, Zr$_4$O, Al$_2$CoO$_4$, Li$_6$V$_3$P$_8$O$_{29}$, ReBi$_3$O$_8$, and LiP(HO$_2$)$_2$. This suggests that experimentally reported or database-documented ``ground truth'' structures may not always represent the thermodynamically most stable phases under the given conditions. These results demonstrate that RL fine-tuning can further improve pretrained generative models via iterative refinement guided by policy gradient-based energy minimization.

\begin{figure}[!t]
  \centering
  \includegraphics[width=\columnwidth]{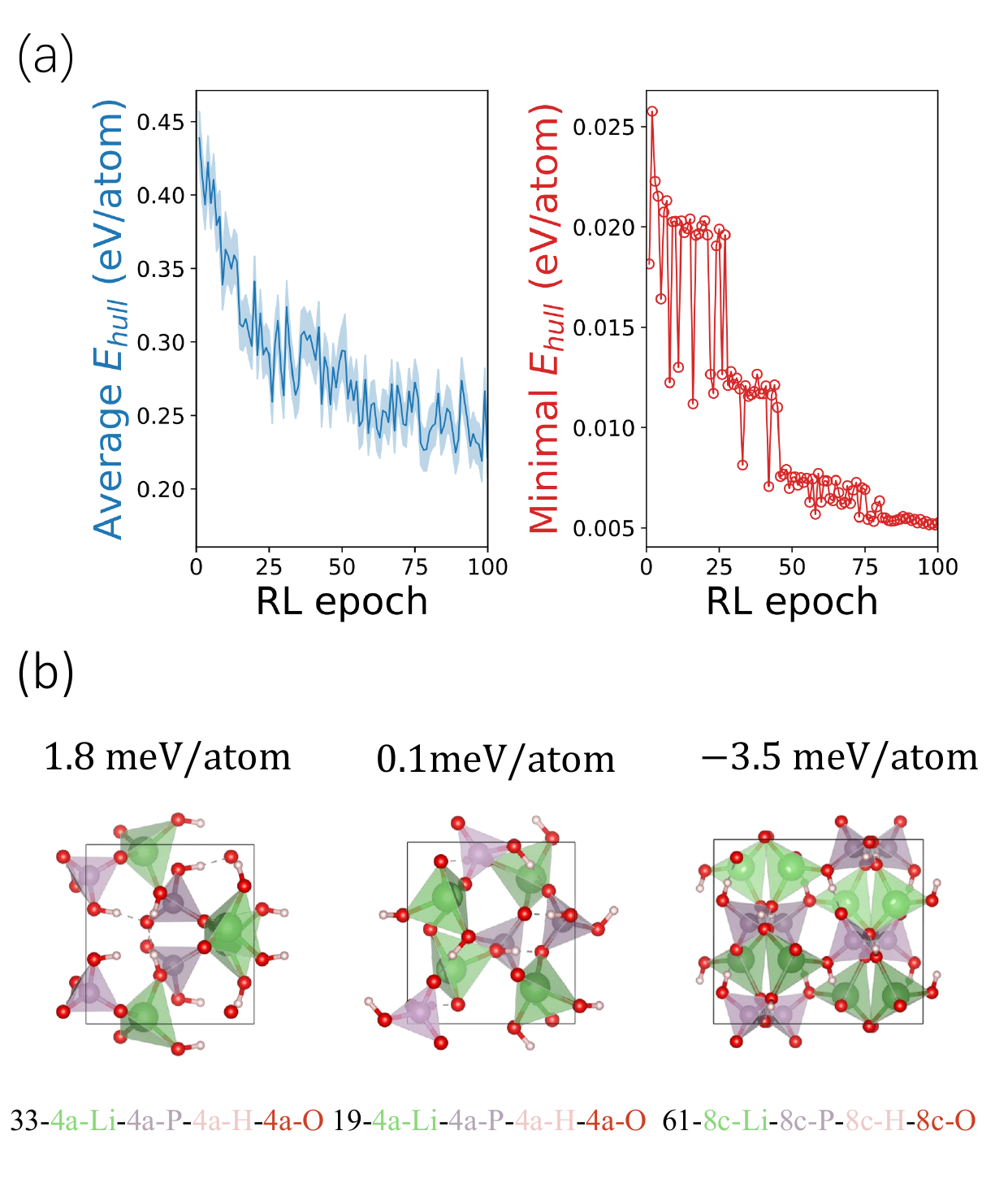}
  \caption{(a) The average and minimal energy above hull of relaxed LiP(HO$_2$)$_2$ versus reinforcement fine-tuning steps. The horizontal line indicates the energy above hull of the ground truth structure. (b) The ground truth structure and discovered structures along their Wyckoff sequences and energy above hull predicted via DFT calculations.
  }
  \label{fig:rl}
\end{figure}

As a concrete demonstration, we present RL fine-tuning results for the LiP(HO$_2$)$_2$ compound. Figure~\ref{fig:rl}(a) shows that average $E_{\textrm{hull}}$ decreases as a function of RL optimization steps. During the RL process, we monitor and plot the lowest energy structures from each batch. After 100 epochs of searching, we identify multiple low-energy configurations with energies similar to the ground truth. Figure~\ref{fig:rl}(b) shows such examples: the first structure matches the ground truth structure in the $Pna2_{1}$ (No. 33) space group, while the second and the third in the $P2_{1}2_{1}2_{1}$ (No. 19) and $Pbca$ (No. 61) space group. These structures exhibit similar structural motifs and show  even lower energies verified via DFT calculations. This example demonstrates that the \csp workflow can discover low-energy polymorphic structures.

\section{Discussion}
\label{sec:discussion}

We present \texttt{CrystalFormer-CSP}, a CSP framework that combines generative pretraining for fast, heuristic structure generation with MLFFs for slow, accurate energy relaxation. We also devise an RL pipeline to fine-tune the generative model using energy above hull evaluated by the MLFF. The approach demonstrates strong performance on benchmark datasets, achieving competitive accuracy with state-of-the-art CSP methods while remaining lightweight and easy to use. We present understandings and list future directions in below.

\subsection{Thinking fast and slow for crystal structure prediction}

The workings of large language models have been widely connected to the dual-process theory of thinking—fast and slow~\cite{kahneman2011thinking}—in cognitive science. Similarly, \csp is a hybrid of System 1 and System 2 thinking for CSP tasks. Generative sampling in CSP can be likened to ``System 1'' thinking: it is fast, heuristic, and relies heavily on empirical patterns learned from large datasets. This stage excels at rapidly proposing candidate structures by leveraging statistical correlations and learned chemical motifs, but may occasionally overlook subtle constraints or generate physically implausible arrangements due to its heuristic nature. In contrast, the subsequent relaxation process functions more like ``System 2'' thinking—slower, deliberate, and grounded in fundamental physical principles such as energy minimization. 

Unlike the sampling stage where all structures are generated with the same computational cost, the relaxation stage allocates computational resources adaptively according to the energy landscape. Depending on the proximity of the initial structure to a local minimum, the relaxation process may take varying numbers of steps. This two-stage pipeline mirrors the dual-process of thinking, where the swift, probabilistic insight of System 1 (generative sampling) is complemented by the careful, principle-driven evaluation of System 2 (relaxation), ultimately leading to the discovery of new crystal structures.

RL provides a mechanism to internalize the energy and geometric insights gained from the ``System 2'' relaxation process back into the ``System 1'' generative model. By fine-tuning the generator using RL with feedback from accurate energy evaluations, the generative model can gradually learn to propose more plausible and lower-energy structures directly, thereby further improving the overall effectiveness of the CSP framework. 

In summary, by unifying heuristic generative modeling with principled energy optimization in a systematically improvable framework, \csp provides a promising new way to the CSP problem. 

\subsection{Future directions}

We expect that with further advances in the following directions, \csp will handle increasingly complex CSP tasks with higher accuracy, reliability, and interpretability. 

\begin{itemize}
    \item \textbf{Force field}: The reliability of energy relaxation and rankings depends on the accuracy of the MLFF. We anticipate that rapid advances in the field of MLFF~\cite{riebesell2025framework} will further improve the accuracy of CSP using the present framework.

    \item \textbf{Finite-temperature and high-pressure structure prediction}: The current method is limited to predicting crystal structures at zero temperature and ambient pressure. The approach could be extended to broader temperature and pressure ranges with appropriate extensions of training data~\cite{opencsp} and force fields~\cite{kruglov2023crystal}. 

    \item \textbf{Organic crystals}: Currently, the model is limited to inorganic crystals due to constraints in training data and model architecture design. It remains unclear whether the same paradigm applies well to organic crystal structure prediction, since space group symmetry inductive bias plays a less significant role in these systems. We expect it may be rewarding to build a system following similar logic based on network topology in reticular chemistry~\cite{o2008reticular} rather than space group symmetries. 

    \item \textbf{Disordered structures}: Refs.~\cite{PRXEnergy.3.011002, cheetham2024artificial} have identified disordered structures as a major challenge for materials generative models. In the present framework, disorder effects from chemical doping could be handled by interpolation of atom feature vectors.

    \item \textbf{Scalability}: The pretrained model fixes the maximum Wyckoff sequence length at 20. Future work should lift this restriction by extending the transformer model's context window.

    \item \textbf{Interpretability}: Applying interpretability techniques to the \csp model would help understand its working mechanism. For example, by generating crystal structures with shared activations, the model may follow element substitution strategies similar to traditional approaches. Interpretability tools could better reveal the chemistry knowledge the model employs for structure predictions.

    \item \textbf{Multi-task pretraining}: Pretraining within the classifier-free guidance framework~\cite{ho2022classifier}, which trains the model both with and without conditioning on chemical formulas, provides a unified approach for both CSP and de novo generation. Future work may explore multi-task learning to further improve performance on both tasks.
  \end{itemize}

In light of these opportunities, we believe that \csp holds great promise to solve CSP challenges in materials discovery and design.

\begin{acknowledgments}
We thank Jian Lv, Han Wang and Kexiang Mao for helpful discussions. This work is supported by the National Natural Science Foundation of China under Grants No. T2225018, No. 92270107, No. 12188101, No. T2121001, the Cross-Disciplinary Key Project of Beijing Natural Science Foundation No. Z250005, the Strategic Priority Research Program of the Chinese Academy of Sciences under Grants No. XDB0500000, and the National Key Projects for Research and Development of China Grants No. 2021YFA1400400. 
\end{acknowledgments}

\bibliography{refs}

%apsrev4-2.bst 2019-01-14 (MD) hand-edited version of apsrev4-1.bst
%Control: key (0)
%Control: author (8) initials jnrlst
%Control: editor formatted (1) identically to author
%Control: production of article title (0) allowed
%Control: page (0) single
%Control: year (1) truncated
%Control: production of eprint (0) enabled
\begin{thebibliography}{50}%
\makeatletter
\providecommand \@ifxundefined [1]{%
 \@ifx{#1\undefined}
}%
\providecommand \@ifnum [1]{%
 \ifnum #1\expandafter \@firstoftwo
 \else \expandafter \@secondoftwo
 \fi
}%
\providecommand \@ifx [1]{%
 \ifx #1\expandafter \@firstoftwo
 \else \expandafter \@secondoftwo
 \fi
}%
\providecommand \natexlab [1]{#1}%
\providecommand \enquote  [1]{``#1''}%
\providecommand \bibnamefont  [1]{#1}%
\providecommand \bibfnamefont [1]{#1}%
\providecommand \citenamefont [1]{#1}%
\providecommand \href@noop [0]{\@secondoftwo}%
\providecommand \href [0]{\begingroup \@sanitize@url \@href}%
\providecommand \@href[1]{\@@startlink{#1}\@@href}%
\providecommand \@@href[1]{\endgroup#1\@@endlink}%
\providecommand \@sanitize@url [0]{\catcode `\\12\catcode `\$12\catcode
  `\&12\catcode `\#12\catcode `\^12\catcode `\_12\catcode `\%12\relax}%
\providecommand \@@startlink[1]{}%
\providecommand \@@endlink[0]{}%
\providecommand \url  [0]{\begingroup\@sanitize@url \@url }%
\providecommand \@url [1]{\endgroup\@href {#1}{\urlprefix }}%
\providecommand \urlprefix  [0]{URL }%
\providecommand \Eprint [0]{\href }%
\providecommand \doibase [0]{https://doi.org/}%
\providecommand \selectlanguage [0]{\@gobble}%
\providecommand \bibinfo  [0]{\@secondoftwo}%
\providecommand \bibfield  [0]{\@secondoftwo}%
\providecommand \translation [1]{[#1]}%
\providecommand \BibitemOpen [0]{}%
\providecommand \bibitemStop [0]{}%
\providecommand \bibitemNoStop [0]{.\EOS\space}%
\providecommand \EOS [0]{\spacefactor3000\relax}%
\providecommand \BibitemShut  [1]{\csname bibitem#1\endcsname}%
\let\auto@bib@innerbib\@empty
%</preamble>
\bibitem [{\citenamefont {Woodley}\ and\ \citenamefont
  {Catlow}(2008)}]{woodley2019crystal}%
  \BibitemOpen
  \bibfield  {author} {\bibinfo {author} {\bibfnamefont {S.}~\bibnamefont
  {Woodley}}\ and\ \bibinfo {author} {\bibfnamefont {R.}~\bibnamefont
  {Catlow}},\ }\bibfield  {title} {\bibinfo {title} {Crystal structure
  prediction from first principles},\ }\href
  {https://www.nature.com/articles/nmat2321} {\bibfield  {journal} {\bibinfo
  {journal} {Nature Materials}\ }\textbf {\bibinfo {volume} {7}},\ \bibinfo
  {pages} {937} (\bibinfo {year} {2008})}\BibitemShut {NoStop}%
\bibitem [{\citenamefont {Oganov}\ \emph {et~al.}(2019)\citenamefont {Oganov},
  \citenamefont {Pickard}, \citenamefont {Zhu},\ and\ \citenamefont
  {Needs}}]{oganov2019structure}%
  \BibitemOpen
  \bibfield  {author} {\bibinfo {author} {\bibfnamefont {A.~R.}\ \bibnamefont
  {Oganov}}, \bibinfo {author} {\bibfnamefont {C.~J.}\ \bibnamefont {Pickard}},
  \bibinfo {author} {\bibfnamefont {Q.}~\bibnamefont {Zhu}},\ and\ \bibinfo
  {author} {\bibfnamefont {R.~J.}\ \bibnamefont {Needs}},\ }\bibfield  {title}
  {\bibinfo {title} {Structure prediction drives materials discovery},\ }\href
  {https://www.nature.com/articles/s41578-019-0101-8} {\bibfield  {journal}
  {\bibinfo  {journal} {Nature Reviews Materials}\ }\textbf {\bibinfo {volume}
  {4}},\ \bibinfo {pages} {331} (\bibinfo {year} {2019})}\BibitemShut {NoStop}%
\bibitem [{\citenamefont {Pauling}(1929)}]{Pauling1929}%
  \BibitemOpen
  \bibfield  {author} {\bibinfo {author} {\bibfnamefont {L.}~\bibnamefont
  {Pauling}},\ }\bibfield  {title} {\bibinfo {title} {The principles
  determining the structure of complex ionic crystals},\ }\href
  {https://doi.org/10.1021/ja01379a006} {\bibfield  {journal} {\bibinfo
  {journal} {Journal of the American Chemical Society}\ }\textbf {\bibinfo
  {volume} {51}},\ \bibinfo {pages} {1010} (\bibinfo {year}
  {1929})}\BibitemShut {NoStop}%
\bibitem [{\citenamefont {Goldschmidt}(1926)}]{Goldschmidt1926}%
  \BibitemOpen
  \bibfield  {author} {\bibinfo {author} {\bibfnamefont {V.~M.}\ \bibnamefont
  {Goldschmidt}},\ }\bibfield  {title} {\bibinfo {title} {Die gesetze der
  krystallochemie},\ }\href {https://doi.org/10.1007/BF01507527} {\bibfield
  {journal} {\bibinfo  {journal} {Naturwissenschaften}\ }\textbf {\bibinfo
  {volume} {14}},\ \bibinfo {pages} {477} (\bibinfo {year} {1926})}\BibitemShut
  {NoStop}%
\bibitem [{\citenamefont {Bartel}\ \emph {et~al.}(2019)\citenamefont {Bartel},
  \citenamefont {Sutton}, \citenamefont {Goldsmith}, \citenamefont {Ouyang},
  \citenamefont {Musgrave}, \citenamefont {Ghiringhelli},\ and\ \citenamefont
  {Scheffler}}]{sciadv.aav0693}%
  \BibitemOpen
  \bibfield  {author} {\bibinfo {author} {\bibfnamefont {C.~J.}\ \bibnamefont
  {Bartel}}, \bibinfo {author} {\bibfnamefont {C.}~\bibnamefont {Sutton}},
  \bibinfo {author} {\bibfnamefont {B.~R.}\ \bibnamefont {Goldsmith}}, \bibinfo
  {author} {\bibfnamefont {R.}~\bibnamefont {Ouyang}}, \bibinfo {author}
  {\bibfnamefont {C.~B.}\ \bibnamefont {Musgrave}}, \bibinfo {author}
  {\bibfnamefont {L.~M.}\ \bibnamefont {Ghiringhelli}},\ and\ \bibinfo {author}
  {\bibfnamefont {M.}~\bibnamefont {Scheffler}},\ }\bibfield  {title} {\bibinfo
  {title} {New tolerance factor to predict the stability of perovskite oxides
  and halides},\ }\href {https://doi.org/10.1126/sciadv.aav0693} {\bibfield
  {journal} {\bibinfo  {journal} {Science Advances}\ }\textbf {\bibinfo
  {volume} {5}},\ \bibinfo {pages} {eaav0693} (\bibinfo {year}
  {2019})}\BibitemShut {NoStop}%
\bibitem [{\citenamefont {George}\ \emph {et~al.}(2020)\citenamefont {George},
  \citenamefont {Waroquiers}, \citenamefont {Di~Stefano}, \citenamefont
  {Petretto}, \citenamefont {Rignanese},\ and\ \citenamefont
  {Hautier}}]{George2020_PaulingRules_Limitations}%
  \BibitemOpen
  \bibfield  {author} {\bibinfo {author} {\bibfnamefont {J.}~\bibnamefont
  {George}}, \bibinfo {author} {\bibfnamefont {D.}~\bibnamefont {Waroquiers}},
  \bibinfo {author} {\bibfnamefont {D.}~\bibnamefont {Di~Stefano}}, \bibinfo
  {author} {\bibfnamefont {G.}~\bibnamefont {Petretto}}, \bibinfo {author}
  {\bibfnamefont {G.}~\bibnamefont {Rignanese}},\ and\ \bibinfo {author}
  {\bibfnamefont {G.}~\bibnamefont {Hautier}},\ }\bibfield  {title} {\bibinfo
  {title} {The limited predictive power of the pauling rules},\ }\href
  {https://doi.org/10.1002/anie.202000829} {\bibfield  {journal} {\bibinfo
  {journal} {Angewandte Chemie International Edition}\ }\textbf {\bibinfo
  {volume} {59}},\ \bibinfo {pages} {7569} (\bibinfo {year}
  {2020})}\BibitemShut {NoStop}%
\bibitem [{\citenamefont {Wang}\ \emph {et~al.}(2012)\citenamefont {Wang},
  \citenamefont {Lv}, \citenamefont {Zhu},\ and\ \citenamefont
  {Ma}}]{wang2012calypso}%
  \BibitemOpen
  \bibfield  {author} {\bibinfo {author} {\bibfnamefont {Y.}~\bibnamefont
  {Wang}}, \bibinfo {author} {\bibfnamefont {J.}~\bibnamefont {Lv}}, \bibinfo
  {author} {\bibfnamefont {L.}~\bibnamefont {Zhu}},\ and\ \bibinfo {author}
  {\bibfnamefont {Y.}~\bibnamefont {Ma}},\ }\bibfield  {title} {\bibinfo
  {title} {Calypso: A method for crystal structure prediction},\ }\href
  {https://doi.org/10.1016/j.cpc.2012.05.008} {\bibfield  {journal} {\bibinfo
  {journal} {Computer Physics Communications}\ }\textbf {\bibinfo {volume}
  {183}},\ \bibinfo {pages} {2063} (\bibinfo {year} {2012})}\BibitemShut
  {NoStop}%
\bibitem [{\citenamefont {Pickard}\ and\ \citenamefont
  {Needs}(2011)}]{pickard2011airss}%
  \BibitemOpen
  \bibfield  {author} {\bibinfo {author} {\bibfnamefont {C.}~\bibnamefont
  {Pickard}}\ and\ \bibinfo {author} {\bibfnamefont {R.}~\bibnamefont
  {Needs}},\ }\bibfield  {title} {\bibinfo {title} {Ab initio random structure
  searching},\ }\href
  {https://iopscience.iop.org/article/10.1088/0953-8984/23/5/053201} {\bibfield
   {journal} {\bibinfo  {journal} {Journal of Physics: Condensed Matter}\
  }\textbf {\bibinfo {volume} {23}},\ \bibinfo {pages} {053201} (\bibinfo
  {year} {2011})}\BibitemShut {NoStop}%
\bibitem [{\citenamefont {Glass}\ \emph {et~al.}(2006)\citenamefont {Glass},
  \citenamefont {Oganov},\ and\ \citenamefont {Hansen}}]{glass2006uspex}%
  \BibitemOpen
  \bibfield  {author} {\bibinfo {author} {\bibfnamefont {C.}~\bibnamefont
  {Glass}}, \bibinfo {author} {\bibfnamefont {A.}~\bibnamefont {Oganov}},\ and\
  \bibinfo {author} {\bibfnamefont {N.}~\bibnamefont {Hansen}},\ }\bibfield
  {title} {\bibinfo {title} {Uspex—evolutionary crystal structure
  prediction},\ }\href
  {https://www.sciencedirect.com/science/article/pii/S0010465506002931}
  {\bibfield  {journal} {\bibinfo  {journal} {Computer Physics Communications}\
  }\textbf {\bibinfo {volume} {175}},\ \bibinfo {pages} {713} (\bibinfo {year}
  {2006})}\BibitemShut {NoStop}%
\bibitem [{\citenamefont {Wang}\ \emph {et~al.}(2023)\citenamefont {Wang},
  \citenamefont {Wang}, \citenamefont {Gao}, \citenamefont {Zhang},
  \citenamefont {Lv}, \citenamefont {Wang}, \citenamefont {Liu}, \citenamefont
  {Wang},\ and\ \citenamefont {Ma}}]{wang2023data}%
  \BibitemOpen
  \bibfield  {author} {\bibinfo {author} {\bibfnamefont {X.}~\bibnamefont
  {Wang}}, \bibinfo {author} {\bibfnamefont {Z.}~\bibnamefont {Wang}}, \bibinfo
  {author} {\bibfnamefont {P.}~\bibnamefont {Gao}}, \bibinfo {author}
  {\bibfnamefont {C.}~\bibnamefont {Zhang}}, \bibinfo {author} {\bibfnamefont
  {J.}~\bibnamefont {Lv}}, \bibinfo {author} {\bibfnamefont {H.}~\bibnamefont
  {Wang}}, \bibinfo {author} {\bibfnamefont {H.}~\bibnamefont {Liu}}, \bibinfo
  {author} {\bibfnamefont {Y.}~\bibnamefont {Wang}},\ and\ \bibinfo {author}
  {\bibfnamefont {Y.}~\bibnamefont {Ma}},\ }\bibfield  {title} {\bibinfo
  {title} {Data-driven prediction of complex crystal structures of dense
  lithium},\ }\href {https://www.nature.com/articles/s41467-023-38650-y}
  {\bibfield  {journal} {\bibinfo  {journal} {Nature Communications}\ }\textbf
  {\bibinfo {volume} {14}},\ \bibinfo {pages} {2924} (\bibinfo {year}
  {2023})}\BibitemShut {NoStop}%
\bibitem [{\citenamefont {Rhodes}\ \emph {et~al.}(2025)\citenamefont {Rhodes},
  \citenamefont {Vandenhaute}, \citenamefont {Šimkus}, \citenamefont {Gin},
  \citenamefont {Godwin}, \citenamefont {Duignan},\ and\ \citenamefont
  {Neumann}}]{orb-v3}%
  \BibitemOpen
  \bibfield  {author} {\bibinfo {author} {\bibfnamefont {B.}~\bibnamefont
  {Rhodes}}, \bibinfo {author} {\bibfnamefont {S.}~\bibnamefont {Vandenhaute}},
  \bibinfo {author} {\bibfnamefont {V.}~\bibnamefont {Šimkus}}, \bibinfo
  {author} {\bibfnamefont {J.}~\bibnamefont {Gin}}, \bibinfo {author}
  {\bibfnamefont {J.}~\bibnamefont {Godwin}}, \bibinfo {author} {\bibfnamefont
  {T.}~\bibnamefont {Duignan}},\ and\ \bibinfo {author} {\bibfnamefont
  {M.}~\bibnamefont {Neumann}},\ }\href {https://arxiv.org/abs/2504.06231}
  {\bibinfo {title} {Orb-v3: atomistic simulation at scale}} (\bibinfo {year}
  {2025}),\ \Eprint {https://arxiv.org/abs/2504.06231} {arXiv:2504.06231
  [cond-mat.mtrl-sci]} \BibitemShut {NoStop}%
\bibitem [{\citenamefont {Zhang}\ \emph {et~al.}(2023)\citenamefont {Zhang},
  \citenamefont {Liu}, \citenamefont {Zhang}, \citenamefont {Zhang},
  \citenamefont {Cai}, \citenamefont {Bi}, \citenamefont {Du}, \citenamefont
  {Qin}, \citenamefont {Huang}, \citenamefont {Li}, \citenamefont {Shan},
  \citenamefont {Zeng}, \citenamefont {Zhang}, \citenamefont {Liu},
  \citenamefont {Li}, \citenamefont {Chang}, \citenamefont {Wang},
  \citenamefont {Zhou}, \citenamefont {Liu}, \citenamefont {Luo}, \citenamefont
  {Wang}, \citenamefont {Jiang}, \citenamefont {Wu}, \citenamefont {Yang},
  \citenamefont {Yang}, \citenamefont {Yang}, \citenamefont {Gong},
  \citenamefont {Zhang}, \citenamefont {Shi}, \citenamefont {Dai},
  \citenamefont {York}, \citenamefont {Liu}, \citenamefont {Zhu}, \citenamefont
  {Zhong}, \citenamefont {Lv}, \citenamefont {Cheng}, \citenamefont {Jia},
  \citenamefont {Chen}, \citenamefont {Ke}, \citenamefont {E}, \citenamefont
  {Zhang},\ and\ \citenamefont {Wang}}]{zhang2023dpa2}%
  \BibitemOpen
  \bibfield  {author} {\bibinfo {author} {\bibfnamefont {D.}~\bibnamefont
  {Zhang}}, \bibinfo {author} {\bibfnamefont {X.}~\bibnamefont {Liu}}, \bibinfo
  {author} {\bibfnamefont {X.}~\bibnamefont {Zhang}}, \bibinfo {author}
  {\bibfnamefont {C.}~\bibnamefont {Zhang}}, \bibinfo {author} {\bibfnamefont
  {C.}~\bibnamefont {Cai}}, \bibinfo {author} {\bibfnamefont {H.}~\bibnamefont
  {Bi}}, \bibinfo {author} {\bibfnamefont {Y.}~\bibnamefont {Du}}, \bibinfo
  {author} {\bibfnamefont {X.}~\bibnamefont {Qin}}, \bibinfo {author}
  {\bibfnamefont {J.}~\bibnamefont {Huang}}, \bibinfo {author} {\bibfnamefont
  {B.}~\bibnamefont {Li}}, \bibinfo {author} {\bibfnamefont {Y.}~\bibnamefont
  {Shan}}, \bibinfo {author} {\bibfnamefont {J.}~\bibnamefont {Zeng}}, \bibinfo
  {author} {\bibfnamefont {Y.}~\bibnamefont {Zhang}}, \bibinfo {author}
  {\bibfnamefont {S.}~\bibnamefont {Liu}}, \bibinfo {author} {\bibfnamefont
  {Y.}~\bibnamefont {Li}}, \bibinfo {author} {\bibfnamefont {J.}~\bibnamefont
  {Chang}}, \bibinfo {author} {\bibfnamefont {X.}~\bibnamefont {Wang}},
  \bibinfo {author} {\bibfnamefont {S.}~\bibnamefont {Zhou}}, \bibinfo {author}
  {\bibfnamefont {J.}~\bibnamefont {Liu}}, \bibinfo {author} {\bibfnamefont
  {X.}~\bibnamefont {Luo}}, \bibinfo {author} {\bibfnamefont {Z.}~\bibnamefont
  {Wang}}, \bibinfo {author} {\bibfnamefont {W.}~\bibnamefont {Jiang}},
  \bibinfo {author} {\bibfnamefont {J.}~\bibnamefont {Wu}}, \bibinfo {author}
  {\bibfnamefont {Y.}~\bibnamefont {Yang}}, \bibinfo {author} {\bibfnamefont
  {J.}~\bibnamefont {Yang}}, \bibinfo {author} {\bibfnamefont {M.}~\bibnamefont
  {Yang}}, \bibinfo {author} {\bibfnamefont {F.-Q.}\ \bibnamefont {Gong}},
  \bibinfo {author} {\bibfnamefont {L.}~\bibnamefont {Zhang}}, \bibinfo
  {author} {\bibfnamefont {M.}~\bibnamefont {Shi}}, \bibinfo {author}
  {\bibfnamefont {F.-Z.}\ \bibnamefont {Dai}}, \bibinfo {author} {\bibfnamefont
  {D.~M.}\ \bibnamefont {York}}, \bibinfo {author} {\bibfnamefont
  {S.}~\bibnamefont {Liu}}, \bibinfo {author} {\bibfnamefont {T.}~\bibnamefont
  {Zhu}}, \bibinfo {author} {\bibfnamefont {Z.}~\bibnamefont {Zhong}}, \bibinfo
  {author} {\bibfnamefont {J.}~\bibnamefont {Lv}}, \bibinfo {author}
  {\bibfnamefont {J.}~\bibnamefont {Cheng}}, \bibinfo {author} {\bibfnamefont
  {W.}~\bibnamefont {Jia}}, \bibinfo {author} {\bibfnamefont {M.}~\bibnamefont
  {Chen}}, \bibinfo {author} {\bibfnamefont {G.}~\bibnamefont {Ke}}, \bibinfo
  {author} {\bibfnamefont {W.}~\bibnamefont {E}}, \bibinfo {author}
  {\bibfnamefont {L.}~\bibnamefont {Zhang}},\ and\ \bibinfo {author}
  {\bibfnamefont {H.}~\bibnamefont {Wang}},\ }\bibfield  {title} {\bibinfo
  {title} {Dpa-2: Towards a universal large atomic model for molecular and
  material simulation},\ }\href@noop {} {\  (\bibinfo {year} {2023})},\ \Eprint
  {https://arxiv.org/abs/2312.15492} {arXiv:2312.15492 [physics.chem-ph]}
  \BibitemShut {NoStop}%
\bibitem [{\citenamefont {Batatia}\ \emph {et~al.}(2024)\citenamefont
  {Batatia}, \citenamefont {Benner}, \citenamefont {Chiang}, \citenamefont
  {Elena}, \citenamefont {Kovács}, \citenamefont {Riebesell}, \citenamefont
  {Advincula}, \citenamefont {Asta}, \citenamefont {Avaylon}, \citenamefont
  {Baldwin}, \citenamefont {Berger}, \citenamefont {Bernstein}, \citenamefont
  {Bhowmik}, \citenamefont {Blau}, \citenamefont {Cărare}, \citenamefont
  {Darby}, \citenamefont {De}, \citenamefont {Pia}, \citenamefont {Deringer},
  \citenamefont {Elijošius}, \citenamefont {El-Machachi}, \citenamefont
  {Falcioni}, \citenamefont {Fako}, \citenamefont {Ferrari}, \citenamefont
  {Genreith-Schriever}, \citenamefont {George}, \citenamefont {Goodall},
  \citenamefont {Grey}, \citenamefont {Grigorev}, \citenamefont {Han},
  \citenamefont {Handley}, \citenamefont {Heenen}, \citenamefont {Hermansson},
  \citenamefont {Holm}, \citenamefont {Jaafar}, \citenamefont {Hofmann},
  \citenamefont {Jakob}, \citenamefont {Jung}, \citenamefont {Kapil},
  \citenamefont {Kaplan}, \citenamefont {Karimitari}, \citenamefont {Kermode},
  \citenamefont {Kroupa}, \citenamefont {Kullgren}, \citenamefont {Kuner},
  \citenamefont {Kuryla}, \citenamefont {Liepuoniute}, \citenamefont {Margraf},
  \citenamefont {Magdău}, \citenamefont {Michaelides}, \citenamefont {Moore},
  \citenamefont {Naik}, \citenamefont {Niblett}, \citenamefont {Norwood},
  \citenamefont {O'Neill}, \citenamefont {Ortner}, \citenamefont {Persson},
  \citenamefont {Reuter}, \citenamefont {Rosen}, \citenamefont {Schaaf},
  \citenamefont {Schran}, \citenamefont {Shi}, \citenamefont {Sivonxay},
  \citenamefont {Stenczel}, \citenamefont {Svahn}, \citenamefont {Sutton},
  \citenamefont {Swinburne}, \citenamefont {Tilly}, \citenamefont {van~der
  Oord}, \citenamefont {Varga-Umbrich}, \citenamefont {Vegge}, \citenamefont
  {Vondrák}, \citenamefont {Wang}, \citenamefont {Witt}, \citenamefont
  {Zills},\ and\ \citenamefont {Csányi}}]{batatia2024foundation}%
  \BibitemOpen
  \bibfield  {author} {\bibinfo {author} {\bibfnamefont {I.}~\bibnamefont
  {Batatia}}, \bibinfo {author} {\bibfnamefont {P.}~\bibnamefont {Benner}},
  \bibinfo {author} {\bibfnamefont {Y.}~\bibnamefont {Chiang}}, \bibinfo
  {author} {\bibfnamefont {A.~M.}\ \bibnamefont {Elena}}, \bibinfo {author}
  {\bibfnamefont {D.~P.}\ \bibnamefont {Kovács}}, \bibinfo {author}
  {\bibfnamefont {J.}~\bibnamefont {Riebesell}}, \bibinfo {author}
  {\bibfnamefont {X.~R.}\ \bibnamefont {Advincula}}, \bibinfo {author}
  {\bibfnamefont {M.}~\bibnamefont {Asta}}, \bibinfo {author} {\bibfnamefont
  {M.}~\bibnamefont {Avaylon}}, \bibinfo {author} {\bibfnamefont {W.~J.}\
  \bibnamefont {Baldwin}}, \bibinfo {author} {\bibfnamefont {F.}~\bibnamefont
  {Berger}}, \bibinfo {author} {\bibfnamefont {N.}~\bibnamefont {Bernstein}},
  \bibinfo {author} {\bibfnamefont {A.}~\bibnamefont {Bhowmik}}, \bibinfo
  {author} {\bibfnamefont {S.~M.}\ \bibnamefont {Blau}}, \bibinfo {author}
  {\bibfnamefont {V.}~\bibnamefont {Cărare}}, \bibinfo {author} {\bibfnamefont
  {J.~P.}\ \bibnamefont {Darby}}, \bibinfo {author} {\bibfnamefont
  {S.}~\bibnamefont {De}}, \bibinfo {author} {\bibfnamefont {F.~D.}\
  \bibnamefont {Pia}}, \bibinfo {author} {\bibfnamefont {V.~L.}\ \bibnamefont
  {Deringer}}, \bibinfo {author} {\bibfnamefont {R.}~\bibnamefont
  {Elijošius}}, \bibinfo {author} {\bibfnamefont {Z.}~\bibnamefont
  {El-Machachi}}, \bibinfo {author} {\bibfnamefont {F.}~\bibnamefont
  {Falcioni}}, \bibinfo {author} {\bibfnamefont {E.}~\bibnamefont {Fako}},
  \bibinfo {author} {\bibfnamefont {A.~C.}\ \bibnamefont {Ferrari}}, \bibinfo
  {author} {\bibfnamefont {A.}~\bibnamefont {Genreith-Schriever}}, \bibinfo
  {author} {\bibfnamefont {J.}~\bibnamefont {George}}, \bibinfo {author}
  {\bibfnamefont {R.~E.~A.}\ \bibnamefont {Goodall}}, \bibinfo {author}
  {\bibfnamefont {C.~P.}\ \bibnamefont {Grey}}, \bibinfo {author}
  {\bibfnamefont {P.}~\bibnamefont {Grigorev}}, \bibinfo {author}
  {\bibfnamefont {S.}~\bibnamefont {Han}}, \bibinfo {author} {\bibfnamefont
  {W.}~\bibnamefont {Handley}}, \bibinfo {author} {\bibfnamefont {H.~H.}\
  \bibnamefont {Heenen}}, \bibinfo {author} {\bibfnamefont {K.}~\bibnamefont
  {Hermansson}}, \bibinfo {author} {\bibfnamefont {C.}~\bibnamefont {Holm}},
  \bibinfo {author} {\bibfnamefont {J.}~\bibnamefont {Jaafar}}, \bibinfo
  {author} {\bibfnamefont {S.}~\bibnamefont {Hofmann}}, \bibinfo {author}
  {\bibfnamefont {K.~S.}\ \bibnamefont {Jakob}}, \bibinfo {author}
  {\bibfnamefont {H.}~\bibnamefont {Jung}}, \bibinfo {author} {\bibfnamefont
  {V.}~\bibnamefont {Kapil}}, \bibinfo {author} {\bibfnamefont {A.~D.}\
  \bibnamefont {Kaplan}}, \bibinfo {author} {\bibfnamefont {N.}~\bibnamefont
  {Karimitari}}, \bibinfo {author} {\bibfnamefont {J.~R.}\ \bibnamefont
  {Kermode}}, \bibinfo {author} {\bibfnamefont {N.}~\bibnamefont {Kroupa}},
  \bibinfo {author} {\bibfnamefont {J.}~\bibnamefont {Kullgren}}, \bibinfo
  {author} {\bibfnamefont {M.~C.}\ \bibnamefont {Kuner}}, \bibinfo {author}
  {\bibfnamefont {D.}~\bibnamefont {Kuryla}}, \bibinfo {author} {\bibfnamefont
  {G.}~\bibnamefont {Liepuoniute}}, \bibinfo {author} {\bibfnamefont {J.~T.}\
  \bibnamefont {Margraf}}, \bibinfo {author} {\bibfnamefont {I.-B.}\
  \bibnamefont {Magdău}}, \bibinfo {author} {\bibfnamefont {A.}~\bibnamefont
  {Michaelides}}, \bibinfo {author} {\bibfnamefont {J.~H.}\ \bibnamefont
  {Moore}}, \bibinfo {author} {\bibfnamefont {A.~A.}\ \bibnamefont {Naik}},
  \bibinfo {author} {\bibfnamefont {S.~P.}\ \bibnamefont {Niblett}}, \bibinfo
  {author} {\bibfnamefont {S.~W.}\ \bibnamefont {Norwood}}, \bibinfo {author}
  {\bibfnamefont {N.}~\bibnamefont {O'Neill}}, \bibinfo {author} {\bibfnamefont
  {C.}~\bibnamefont {Ortner}}, \bibinfo {author} {\bibfnamefont {K.~A.}\
  \bibnamefont {Persson}}, \bibinfo {author} {\bibfnamefont {K.}~\bibnamefont
  {Reuter}}, \bibinfo {author} {\bibfnamefont {A.~S.}\ \bibnamefont {Rosen}},
  \bibinfo {author} {\bibfnamefont {L.~L.}\ \bibnamefont {Schaaf}}, \bibinfo
  {author} {\bibfnamefont {C.}~\bibnamefont {Schran}}, \bibinfo {author}
  {\bibfnamefont {B.~X.}\ \bibnamefont {Shi}}, \bibinfo {author} {\bibfnamefont
  {E.}~\bibnamefont {Sivonxay}}, \bibinfo {author} {\bibfnamefont {T.~K.}\
  \bibnamefont {Stenczel}}, \bibinfo {author} {\bibfnamefont {V.}~\bibnamefont
  {Svahn}}, \bibinfo {author} {\bibfnamefont {C.}~\bibnamefont {Sutton}},
  \bibinfo {author} {\bibfnamefont {T.~D.}\ \bibnamefont {Swinburne}}, \bibinfo
  {author} {\bibfnamefont {J.}~\bibnamefont {Tilly}}, \bibinfo {author}
  {\bibfnamefont {C.}~\bibnamefont {van~der Oord}}, \bibinfo {author}
  {\bibfnamefont {E.}~\bibnamefont {Varga-Umbrich}}, \bibinfo {author}
  {\bibfnamefont {T.}~\bibnamefont {Vegge}}, \bibinfo {author} {\bibfnamefont
  {M.}~\bibnamefont {Vondrák}}, \bibinfo {author} {\bibfnamefont
  {Y.}~\bibnamefont {Wang}}, \bibinfo {author} {\bibfnamefont {W.~C.}\
  \bibnamefont {Witt}}, \bibinfo {author} {\bibfnamefont {F.}~\bibnamefont
  {Zills}},\ and\ \bibinfo {author} {\bibfnamefont {G.}~\bibnamefont
  {Csányi}},\ }\bibfield  {title} {\bibinfo {title} {A foundation model for
  atomistic materials chemistry},\ }\href@noop {} {\  (\bibinfo {year}
  {2024})},\ \Eprint {https://arxiv.org/abs/2401.00096} {arXiv:2401.00096
  [physics.chem-ph]} \BibitemShut {NoStop}%
\bibitem [{\citenamefont {Zhang}\ \emph {et~al.}(2025)\citenamefont {Zhang},
  \citenamefont {Peng}, \citenamefont {Cai}, \citenamefont {Li}, \citenamefont
  {Zhou}, \citenamefont {Zeng}, \citenamefont {Guo}, \citenamefont {Zhang},
  \citenamefont {Li}, \citenamefont {Jiang} \emph {et~al.}}]{zhang2025graph}%
  \BibitemOpen
  \bibfield  {author} {\bibinfo {author} {\bibfnamefont {D.}~\bibnamefont
  {Zhang}}, \bibinfo {author} {\bibfnamefont {A.}~\bibnamefont {Peng}},
  \bibinfo {author} {\bibfnamefont {C.}~\bibnamefont {Cai}}, \bibinfo {author}
  {\bibfnamefont {W.}~\bibnamefont {Li}}, \bibinfo {author} {\bibfnamefont
  {Y.}~\bibnamefont {Zhou}}, \bibinfo {author} {\bibfnamefont {J.}~\bibnamefont
  {Zeng}}, \bibinfo {author} {\bibfnamefont {M.}~\bibnamefont {Guo}}, \bibinfo
  {author} {\bibfnamefont {C.}~\bibnamefont {Zhang}}, \bibinfo {author}
  {\bibfnamefont {B.}~\bibnamefont {Li}}, \bibinfo {author} {\bibfnamefont
  {H.}~\bibnamefont {Jiang}}, \emph {et~al.},\ }\bibfield  {title} {\bibinfo
  {title} {Graph neural network model for the era of large atomistic models},\
  }\href@noop {} {\  (\bibinfo {year} {2025})},\ \Eprint
  {https://arxiv.org/abs/2506.01686} {arXiv:2506.01686 [physics.comp-ph]}
  \BibitemShut {NoStop}%
\bibitem [{\citenamefont {Yang}\ \emph {et~al.}(2024)\citenamefont {Yang},
  \citenamefont {Hu}, \citenamefont {Zhou}, \citenamefont {Liu}, \citenamefont
  {Shi}, \citenamefont {Li}, \citenamefont {Li}, \citenamefont {Chen},
  \citenamefont {Chen}, \citenamefont {Zeni} \emph
  {et~al.}}]{yang2024mattersim}%
  \BibitemOpen
  \bibfield  {author} {\bibinfo {author} {\bibfnamefont {H.}~\bibnamefont
  {Yang}}, \bibinfo {author} {\bibfnamefont {C.}~\bibnamefont {Hu}}, \bibinfo
  {author} {\bibfnamefont {Y.}~\bibnamefont {Zhou}}, \bibinfo {author}
  {\bibfnamefont {X.}~\bibnamefont {Liu}}, \bibinfo {author} {\bibfnamefont
  {Y.}~\bibnamefont {Shi}}, \bibinfo {author} {\bibfnamefont {J.}~\bibnamefont
  {Li}}, \bibinfo {author} {\bibfnamefont {G.}~\bibnamefont {Li}}, \bibinfo
  {author} {\bibfnamefont {Z.}~\bibnamefont {Chen}}, \bibinfo {author}
  {\bibfnamefont {S.}~\bibnamefont {Chen}}, \bibinfo {author} {\bibfnamefont
  {C.}~\bibnamefont {Zeni}}, \emph {et~al.},\ }\bibfield  {title} {\bibinfo
  {title} {Mattersim: A deep learning atomistic model across elements,
  temperatures and pressures},\ }\href@noop {} {\  (\bibinfo {year} {2024})},\
  \Eprint {https://arxiv.org/abs/2405.04967} {arXiv:2405.04967
  [cond-mat.mtrl-sci]} \BibitemShut {NoStop}%
\bibitem [{\citenamefont {Kim}\ \emph {et~al.}(2024)\citenamefont {Kim},
  \citenamefont {Kim}, \citenamefont {Kim}, \citenamefont {Lee}, \citenamefont
  {Park}, \citenamefont {Kang},\ and\ \citenamefont {Han}}]{kim2024data}%
  \BibitemOpen
  \bibfield  {author} {\bibinfo {author} {\bibfnamefont {J.}~\bibnamefont
  {Kim}}, \bibinfo {author} {\bibfnamefont {J.}~\bibnamefont {Kim}}, \bibinfo
  {author} {\bibfnamefont {J.}~\bibnamefont {Kim}}, \bibinfo {author}
  {\bibfnamefont {J.}~\bibnamefont {Lee}}, \bibinfo {author} {\bibfnamefont
  {Y.}~\bibnamefont {Park}}, \bibinfo {author} {\bibfnamefont {Y.}~\bibnamefont
  {Kang}},\ and\ \bibinfo {author} {\bibfnamefont {S.}~\bibnamefont {Han}},\
  }\bibfield  {title} {\bibinfo {title} {Data-efficient multifidelity training
  for high-fidelity machine learning interatomic potentials},\ }\href
  {https://pubs.acs.org/doi/10.1021/jacs.4c14455} {\bibfield  {journal}
  {\bibinfo  {journal} {Journal of the American Chemical Society}\ }\textbf
  {\bibinfo {volume} {147}},\ \bibinfo {pages} {1042} (\bibinfo {year}
  {2024})}\BibitemShut {NoStop}%
\bibitem [{\citenamefont {Merchant}\ \emph {et~al.}(2023)\citenamefont
  {Merchant}, \citenamefont {Batzner}, \citenamefont {Schoenholz},
  \citenamefont {Aykol}, \citenamefont {Cheon},\ and\ \citenamefont
  {Cubuk}}]{merchant2023scaling}%
  \BibitemOpen
  \bibfield  {author} {\bibinfo {author} {\bibfnamefont {A.}~\bibnamefont
  {Merchant}}, \bibinfo {author} {\bibfnamefont {S.}~\bibnamefont {Batzner}},
  \bibinfo {author} {\bibfnamefont {S.~S.}\ \bibnamefont {Schoenholz}},
  \bibinfo {author} {\bibfnamefont {M.}~\bibnamefont {Aykol}}, \bibinfo
  {author} {\bibfnamefont {G.}~\bibnamefont {Cheon}},\ and\ \bibinfo {author}
  {\bibfnamefont {E.~D.}\ \bibnamefont {Cubuk}},\ }\bibfield  {title} {\bibinfo
  {title} {Scaling deep learning for materials discovery},\ }\href
  {https://doi.org/10.1038/s41586-023-06735-9} {\bibfield  {journal} {\bibinfo
  {journal} {Nature}\ }\textbf {\bibinfo {volume} {624}},\ \bibinfo {pages}
  {80} (\bibinfo {year} {2023})}\BibitemShut {NoStop}%
\bibitem [{\citenamefont {Schmidt}\ \emph
  {et~al.}(2024{\natexlab{a}})\citenamefont {Schmidt}, \citenamefont
  {Cerqueira}, \citenamefont {Romero}, \citenamefont {Loew}, \citenamefont
  {Jäger}, \citenamefont {Wang}, \citenamefont {Botti},\ and\ \citenamefont
  {Marques}}]{schmidt2024improving}%
  \BibitemOpen
  \bibfield  {author} {\bibinfo {author} {\bibfnamefont {J.}~\bibnamefont
  {Schmidt}}, \bibinfo {author} {\bibfnamefont {T.~F.}\ \bibnamefont
  {Cerqueira}}, \bibinfo {author} {\bibfnamefont {A.~H.}\ \bibnamefont
  {Romero}}, \bibinfo {author} {\bibfnamefont {A.}~\bibnamefont {Loew}},
  \bibinfo {author} {\bibfnamefont {F.}~\bibnamefont {Jäger}}, \bibinfo
  {author} {\bibfnamefont {H.-C.}\ \bibnamefont {Wang}}, \bibinfo {author}
  {\bibfnamefont {S.}~\bibnamefont {Botti}},\ and\ \bibinfo {author}
  {\bibfnamefont {M.~A.}\ \bibnamefont {Marques}},\ }\bibfield  {title}
  {\bibinfo {title} {Improving machine-learning models in materials science
  through large datasets},\ }\href
  {https://doi.org/https://doi.org/10.1016/j.mtphys.2024.101560} {\bibfield
  {journal} {\bibinfo  {journal} {Mater. Today Phys.}\ }\textbf {\bibinfo
  {volume} {48}},\ \bibinfo {pages} {101560} (\bibinfo {year}
  {2024}{\natexlab{a}})}\BibitemShut {NoStop}%
\bibitem [{\citenamefont {Wang}\ \emph
  {et~al.}(2025{\natexlab{a}})\citenamefont {Wang}, \citenamefont {Zhang},
  \citenamefont {Wang}, \citenamefont {Liu}, \citenamefont {Lv}, \citenamefont
  {Wang}, \citenamefont {Ma} \emph {et~al.}}]{wang2025discovery}%
  \BibitemOpen
  \bibfield  {author} {\bibinfo {author} {\bibfnamefont {X.}~\bibnamefont
  {Wang}}, \bibinfo {author} {\bibfnamefont {C.}~\bibnamefont {Zhang}},
  \bibinfo {author} {\bibfnamefont {Z.}~\bibnamefont {Wang}}, \bibinfo {author}
  {\bibfnamefont {H.}~\bibnamefont {Liu}}, \bibinfo {author} {\bibfnamefont
  {J.}~\bibnamefont {Lv}}, \bibinfo {author} {\bibfnamefont {H.}~\bibnamefont
  {Wang}}, \bibinfo {author} {\bibfnamefont {Y.}~\bibnamefont {Ma}}, \emph
  {et~al.},\ }\href@noop {} {\bibinfo {title} {Discovery of high-temperature
  superconducting ternary hydrides via deep learning}} (\bibinfo {year}
  {2025}{\natexlab{a}}),\ \Eprint {https://arxiv.org/abs/2502.16558}
  {arXiv:2502.16558 [cond-mat.mtrl-sci]} \BibitemShut {NoStop}%
\bibitem [{\citenamefont {Li}\ \emph {et~al.}(2025)\citenamefont {Li},
  \citenamefont {Feng}, \citenamefont {Luo}, \citenamefont {Jiang},
  \citenamefont {Zheng}, \citenamefont {Song}, \citenamefont {Lv},
  \citenamefont {Butler}, \citenamefont {Liu}, \citenamefont {Xie},
  \citenamefont {Xie},\ and\ \citenamefont
  {Ma}}]{li2025selfoptimizingmachinelearningpotential}%
  \BibitemOpen
  \bibfield  {author} {\bibinfo {author} {\bibfnamefont {J.}~\bibnamefont
  {Li}}, \bibinfo {author} {\bibfnamefont {J.}~\bibnamefont {Feng}}, \bibinfo
  {author} {\bibfnamefont {J.}~\bibnamefont {Luo}}, \bibinfo {author}
  {\bibfnamefont {B.}~\bibnamefont {Jiang}}, \bibinfo {author} {\bibfnamefont
  {X.}~\bibnamefont {Zheng}}, \bibinfo {author} {\bibfnamefont
  {Q.}~\bibnamefont {Song}}, \bibinfo {author} {\bibfnamefont {J.}~\bibnamefont
  {Lv}}, \bibinfo {author} {\bibfnamefont {K.}~\bibnamefont {Butler}}, \bibinfo
  {author} {\bibfnamefont {H.}~\bibnamefont {Liu}}, \bibinfo {author}
  {\bibfnamefont {C.}~\bibnamefont {Xie}}, \bibinfo {author} {\bibfnamefont
  {Y.}~\bibnamefont {Xie}},\ and\ \bibinfo {author} {\bibfnamefont
  {Y.}~\bibnamefont {Ma}},\ }\href@noop {} {\bibinfo {title} {Self-optimizing
  machine learning potential assisted automated workflow for highly efficient
  complex systems material design}} (\bibinfo {year} {2025}),\ \Eprint
  {https://arxiv.org/abs/2505.08159} {arXiv:2505.08159 [cond-mat.mtrl-sci]}
  \BibitemShut {NoStop}%
\bibitem [{\citenamefont {Xie}\ \emph {et~al.}(2022)\citenamefont {Xie},
  \citenamefont {Fu}, \citenamefont {Ganea}, \citenamefont {Barzilay},\ and\
  \citenamefont {Jaakkola}}]{cdvae}%
  \BibitemOpen
  \bibfield  {author} {\bibinfo {author} {\bibfnamefont {T.}~\bibnamefont
  {Xie}}, \bibinfo {author} {\bibfnamefont {X.}~\bibnamefont {Fu}}, \bibinfo
  {author} {\bibfnamefont {O.-E.}\ \bibnamefont {Ganea}}, \bibinfo {author}
  {\bibfnamefont {R.}~\bibnamefont {Barzilay}},\ and\ \bibinfo {author}
  {\bibfnamefont {T.~S.}\ \bibnamefont {Jaakkola}},\ }\bibfield  {title}
  {\bibinfo {title} {Crystal diffusion variational autoencoder for periodic
  material generation},\ }in\ \href {https://openreview.net/pdf?id=03RLpj-tc_}
  {\emph {\bibinfo {booktitle} {International Conference on Learning
  Representations (ICLR) 2022}}}\ (\bibinfo {year} {2022})\ \bibinfo {note}
  {published as a conference paper at ICLR 2022}\BibitemShut {NoStop}%
\bibitem [{\citenamefont {Jiao}\ \emph {et~al.}(2023)\citenamefont {Jiao},
  \citenamefont {Huang}, \citenamefont {Liu}, \citenamefont {Zhao},\ and\
  \citenamefont {Liu}}]{diffcsp}%
  \BibitemOpen
  \bibfield  {author} {\bibinfo {author} {\bibfnamefont {R.}~\bibnamefont
  {Jiao}}, \bibinfo {author} {\bibfnamefont {W.}~\bibnamefont {Huang}},
  \bibinfo {author} {\bibfnamefont {Y.}~\bibnamefont {Liu}}, \bibinfo {author}
  {\bibfnamefont {D.}~\bibnamefont {Zhao}},\ and\ \bibinfo {author}
  {\bibfnamefont {Y.}~\bibnamefont {Liu}},\ }\bibfield  {title} {\bibinfo
  {title} {Crystal structure prediction by joint equivariant diffusion},\
  }\href
  {https://proceedings.neurips.cc/paper_files/paper/2023/file/38b787fc530d0b31825827e2cc306656-Paper-Conference.pdf}
  {\bibfield  {journal} {\bibinfo  {journal} {Advances in Neural Information
  Processing Systems (NeurIPS)}\ } (\bibinfo {year} {2023})},\ \bibinfo {note}
  {pDF available}\BibitemShut {NoStop}%
\bibitem [{\citenamefont {Zeni}\ \emph {et~al.}(2025)\citenamefont {Zeni},
  \citenamefont {Pinsler}, \citenamefont {Zügner} \emph {et~al.}}]{mattergen}%
  \BibitemOpen
  \bibfield  {author} {\bibinfo {author} {\bibfnamefont {C.}~\bibnamefont
  {Zeni}}, \bibinfo {author} {\bibfnamefont {R.}~\bibnamefont {Pinsler}},
  \bibinfo {author} {\bibfnamefont {D.}~\bibnamefont {Zügner}}, \emph
  {et~al.},\ }\bibfield  {title} {\bibinfo {title} {A generative model for
  inorganic materials design},\ }\href
  {https://doi.org/10.1038/s41586-025-08628-5} {\bibfield  {journal} {\bibinfo
  {journal} {Nature}\ }\textbf {\bibinfo {volume} {639}},\ \bibinfo {pages}
  {624–632} (\bibinfo {year} {2025})}\BibitemShut {NoStop}%
\bibitem [{\citenamefont {Miller}\ \emph {et~al.}(2024)\citenamefont {Miller},
  \citenamefont {Chen}, \citenamefont {Sriram},\ and\ \citenamefont
  {Wood}}]{flowmm}%
  \BibitemOpen
  \bibfield  {author} {\bibinfo {author} {\bibfnamefont {B.~K.}\ \bibnamefont
  {Miller}}, \bibinfo {author} {\bibfnamefont {R.~T.~Q.}\ \bibnamefont {Chen}},
  \bibinfo {author} {\bibfnamefont {A.}~\bibnamefont {Sriram}},\ and\ \bibinfo
  {author} {\bibfnamefont {B.~M.}\ \bibnamefont {Wood}},\ }\bibfield  {title}
  {\bibinfo {title} {Flowmm: Generating materials with riemannian flow
  matching},\ }in\ \href {https://openreview.net/forum?id=W4pB7VbzZI} {\emph
  {\bibinfo {booktitle} {Proceedings of the Forty-first International
  Conference on Machine Learning (ICML 2024)}}}\ (\bibinfo {year}
  {2024})\BibitemShut {NoStop}%
\bibitem [{\citenamefont {Cao}\ \emph {et~al.}(2025)\citenamefont {Cao},
  \citenamefont {Luo}, \citenamefont {Lv},\ and\ \citenamefont
  {Wang}}]{crystalformer}%
  \BibitemOpen
  \bibfield  {author} {\bibinfo {author} {\bibfnamefont {Z.}~\bibnamefont
  {Cao}}, \bibinfo {author} {\bibfnamefont {X.}~\bibnamefont {Luo}}, \bibinfo
  {author} {\bibfnamefont {J.}~\bibnamefont {Lv}},\ and\ \bibinfo {author}
  {\bibfnamefont {L.}~\bibnamefont {Wang}},\ }\bibfield  {title} {\bibinfo
  {title} {Space group informed transformer for crystalline materials
  generation},\ }\href
  {https://doi.org/https://doi.org/10.1016/j.scib.2025.09.035} {\bibfield
  {journal} {\bibinfo  {journal} {Science Bulletin}\ }\textbf {\bibinfo
  {volume} {70}},\ \bibinfo {pages} {3522} (\bibinfo {year}
  {2025})}\BibitemShut {NoStop}%
\bibitem [{\citenamefont {Luo}\ \emph {et~al.}(2025)\citenamefont {Luo},
  \citenamefont {Wang}, \citenamefont {Wang}, \citenamefont {Shao},
  \citenamefont {Lv}, \citenamefont {Wang}, \citenamefont {Wang},\ and\
  \citenamefont {Ma}}]{crystalflow}%
  \BibitemOpen
  \bibfield  {author} {\bibinfo {author} {\bibfnamefont {X.}~\bibnamefont
  {Luo}}, \bibinfo {author} {\bibfnamefont {Z.}~\bibnamefont {Wang}}, \bibinfo
  {author} {\bibfnamefont {Q.}~\bibnamefont {Wang}}, \bibinfo {author}
  {\bibfnamefont {X.}~\bibnamefont {Shao}}, \bibinfo {author} {\bibfnamefont
  {J.}~\bibnamefont {Lv}}, \bibinfo {author} {\bibfnamefont {L.}~\bibnamefont
  {Wang}}, \bibinfo {author} {\bibfnamefont {Y.}~\bibnamefont {Wang}},\ and\
  \bibinfo {author} {\bibfnamefont {Y.}~\bibnamefont {Ma}},\ }\bibfield
  {title} {\bibinfo {title} {Crystalflow: a flow-based generative model for
  crystalline materials},\ }\bibfield  {journal} {\bibinfo  {journal} {Nature
  Communications}\ }\textbf {\bibinfo {volume} {16}},\ \href
  {https://doi.org/10.1038/s41467-025-64364-4} {10.1038/s41467-025-64364-4}
  (\bibinfo {year} {2025})\BibitemShut {NoStop}%
\bibitem [{\citenamefont {Lin}\ \emph {et~al.}(2025)\citenamefont {Lin},
  \citenamefont {Chen}, \citenamefont {Jiao}, \citenamefont {Mo}, \citenamefont
  {Cen}, \citenamefont {Huang}, \citenamefont {Liu}, \citenamefont {Huang},\
  and\ \citenamefont {Lu}}]{equicsp}%
  \BibitemOpen
  \bibfield  {author} {\bibinfo {author} {\bibfnamefont {P.}~\bibnamefont
  {Lin}}, \bibinfo {author} {\bibfnamefont {P.}~\bibnamefont {Chen}}, \bibinfo
  {author} {\bibfnamefont {R.}~\bibnamefont {Jiao}}, \bibinfo {author}
  {\bibfnamefont {Q.}~\bibnamefont {Mo}}, \bibinfo {author} {\bibfnamefont
  {J.}~\bibnamefont {Cen}}, \bibinfo {author} {\bibfnamefont {W.}~\bibnamefont
  {Huang}}, \bibinfo {author} {\bibfnamefont {Y.}~\bibnamefont {Liu}}, \bibinfo
  {author} {\bibfnamefont {D.}~\bibnamefont {Huang}},\ and\ \bibinfo {author}
  {\bibfnamefont {Y.}~\bibnamefont {Lu}},\ }\bibfield  {title} {\bibinfo
  {title} {Equivariant diffusion for crystal structure prediction},\ }\href
  {https://arxiv.org/abs/2512.07289} {\  (\bibinfo {year} {2025})},\ \Eprint
  {https://arxiv.org/abs/2512.07289} {arXiv:2512.07289 [cond-mat.mtrl-sci]}
  \BibitemShut {NoStop}%
\bibitem [{\citenamefont {Schmidt}\ \emph
  {et~al.}(2024{\natexlab{b}})\citenamefont {Schmidt}, \citenamefont
  {Cerqueira}, \citenamefont {Romero}, \citenamefont {Loew}, \citenamefont
  {Jäger}, \citenamefont {Wang}, \citenamefont {Botti},\ and\ \citenamefont
  {Marques}}]{alexandria}%
  \BibitemOpen
  \bibfield  {author} {\bibinfo {author} {\bibfnamefont {J.}~\bibnamefont
  {Schmidt}}, \bibinfo {author} {\bibfnamefont {T.~F.}\ \bibnamefont
  {Cerqueira}}, \bibinfo {author} {\bibfnamefont {A.~H.}\ \bibnamefont
  {Romero}}, \bibinfo {author} {\bibfnamefont {A.}~\bibnamefont {Loew}},
  \bibinfo {author} {\bibfnamefont {F.}~\bibnamefont {Jäger}}, \bibinfo
  {author} {\bibfnamefont {H.-C.}\ \bibnamefont {Wang}}, \bibinfo {author}
  {\bibfnamefont {S.}~\bibnamefont {Botti}},\ and\ \bibinfo {author}
  {\bibfnamefont {M.~A.}\ \bibnamefont {Marques}},\ }\bibfield  {title}
  {\bibinfo {title} {Improving machine-learning models in materials science
  through large datasets},\ }\href
  {https://doi.org/https://doi.org/10.1016/j.mtphys.2024.101560} {\bibfield
  {journal} {\bibinfo  {journal} {Materials Today Physics}\ }\textbf {\bibinfo
  {volume} {48}},\ \bibinfo {pages} {101560} (\bibinfo {year}
  {2024}{\natexlab{b}})}\BibitemShut {NoStop}%
\bibitem [{\citenamefont {Cavignac}\ \emph {et~al.}(2025)\citenamefont
  {Cavignac}, \citenamefont {Schmidt}, \citenamefont {Breuck}, \citenamefont
  {Loew}, \citenamefont {Cerqueira}, \citenamefont {Wang}, \citenamefont
  {Bochkarev}, \citenamefont {Lysogorskiy}, \citenamefont {Romero},
  \citenamefont {Drautz}, \citenamefont {Botti},\ and\ \citenamefont
  {Marques}}]{cavignac2025aidriven}%
  \BibitemOpen
  \bibfield  {author} {\bibinfo {author} {\bibfnamefont {T.}~\bibnamefont
  {Cavignac}}, \bibinfo {author} {\bibfnamefont {J.}~\bibnamefont {Schmidt}},
  \bibinfo {author} {\bibfnamefont {P.-P.~D.}\ \bibnamefont {Breuck}}, \bibinfo
  {author} {\bibfnamefont {A.}~\bibnamefont {Loew}}, \bibinfo {author}
  {\bibfnamefont {T.~F.~T.}\ \bibnamefont {Cerqueira}}, \bibinfo {author}
  {\bibfnamefont {H.-C.}\ \bibnamefont {Wang}}, \bibinfo {author}
  {\bibfnamefont {A.}~\bibnamefont {Bochkarev}}, \bibinfo {author}
  {\bibfnamefont {Y.}~\bibnamefont {Lysogorskiy}}, \bibinfo {author}
  {\bibfnamefont {A.~H.}\ \bibnamefont {Romero}}, \bibinfo {author}
  {\bibfnamefont {R.}~\bibnamefont {Drautz}}, \bibinfo {author} {\bibfnamefont
  {S.}~\bibnamefont {Botti}},\ and\ \bibinfo {author} {\bibfnamefont
  {M.~A.~L.}\ \bibnamefont {Marques}},\ }\bibfield  {title} {\bibinfo {title}
  {Ai-driven expansion and application of the alexandria database},\ }\href
  {https://arxiv.org/abs/2512.09169} {\  (\bibinfo {year} {2025})},\ \Eprint
  {https://arxiv.org/abs/2512.09169} {arXiv:2512.09169 [cond-mat.mtrl-sci]}
  \BibitemShut {NoStop}%
\bibitem [{\citenamefont {Hornfeck}(2022)}]{hornfeck2022combinatorics}%
  \BibitemOpen
  \bibfield  {author} {\bibinfo {author} {\bibfnamefont {W.}~\bibnamefont
  {Hornfeck}},\ }\bibfield  {title} {\bibinfo {title} {On the combinatorics of
  crystal structures: number of wyckoff sequences of given length},\ }\href
  {https://onlinelibrary.wiley.com/iucr/doi/10.1107/S2053273321013565}
  {\bibfield  {journal} {\bibinfo  {journal} {Foundations of Crystallography}\
  }\textbf {\bibinfo {volume} {78}},\ \bibinfo {pages} {149} (\bibinfo {year}
  {2022})}\BibitemShut {NoStop}%
\bibitem [{\citenamefont {Delétang}\ \emph {et~al.}(2024)\citenamefont
  {Delétang}, \citenamefont {Ruoss}, \citenamefont {Duquenne}, \citenamefont
  {Catt}, \citenamefont {Genewein}, \citenamefont {Mattern}, \citenamefont
  {Grau-Moya}, \citenamefont {Wenliang}, \citenamefont {Aitchison},
  \citenamefont {Orseau}, \citenamefont {Hutter},\ and\ \citenamefont
  {Veness}}]{Deletang2024_LMIsCompression}%
  \BibitemOpen
  \bibfield  {author} {\bibinfo {author} {\bibfnamefont {G.}~\bibnamefont
  {Delétang}}, \bibinfo {author} {\bibfnamefont {A.}~\bibnamefont {Ruoss}},
  \bibinfo {author} {\bibfnamefont {P.-A.}\ \bibnamefont {Duquenne}}, \bibinfo
  {author} {\bibfnamefont {E.}~\bibnamefont {Catt}}, \bibinfo {author}
  {\bibfnamefont {T.}~\bibnamefont {Genewein}}, \bibinfo {author}
  {\bibfnamefont {C.}~\bibnamefont {Mattern}}, \bibinfo {author} {\bibfnamefont
  {J.}~\bibnamefont {Grau-Moya}}, \bibinfo {author} {\bibfnamefont {L.~K.}\
  \bibnamefont {Wenliang}}, \bibinfo {author} {\bibfnamefont {M.}~\bibnamefont
  {Aitchison}}, \bibinfo {author} {\bibfnamefont {L.}~\bibnamefont {Orseau}},
  \bibinfo {author} {\bibfnamefont {M.}~\bibnamefont {Hutter}},\ and\ \bibinfo
  {author} {\bibfnamefont {J.}~\bibnamefont {Veness}},\ }\href@noop {}
  {\bibinfo {title} {Language modeling is compression}} (\bibinfo {year}
  {2024}),\ \Eprint {https://arxiv.org/abs/2309.10668} {arXiv:2309.10668
  [cs.LG]} \BibitemShut {NoStop}%
\bibitem [{\citenamefont {Riebesell}\ \emph {et~al.}(2025)\citenamefont
  {Riebesell}, \citenamefont {Goodall}, \citenamefont {Benner}, \citenamefont
  {Chiang}, \citenamefont {Deng}, \citenamefont {Ceder}, \citenamefont {Asta},
  \citenamefont {Lee}, \citenamefont {Jain},\ and\ \citenamefont
  {Persson}}]{riebesell2025framework}%
  \BibitemOpen
  \bibfield  {author} {\bibinfo {author} {\bibfnamefont {J.}~\bibnamefont
  {Riebesell}}, \bibinfo {author} {\bibfnamefont {R.~E.}\ \bibnamefont
  {Goodall}}, \bibinfo {author} {\bibfnamefont {P.}~\bibnamefont {Benner}},
  \bibinfo {author} {\bibfnamefont {Y.}~\bibnamefont {Chiang}}, \bibinfo
  {author} {\bibfnamefont {B.}~\bibnamefont {Deng}}, \bibinfo {author}
  {\bibfnamefont {G.}~\bibnamefont {Ceder}}, \bibinfo {author} {\bibfnamefont
  {M.}~\bibnamefont {Asta}}, \bibinfo {author} {\bibfnamefont {A.~A.}\
  \bibnamefont {Lee}}, \bibinfo {author} {\bibfnamefont {A.}~\bibnamefont
  {Jain}},\ and\ \bibinfo {author} {\bibfnamefont {K.~A.}\ \bibnamefont
  {Persson}},\ }\bibfield  {title} {\bibinfo {title} {A framework to evaluate
  machine learning crystal stability predictions},\ }\href
  {https://www.nature.com/articles/s42256-025-01055-1} {\bibfield  {journal}
  {\bibinfo  {journal} {Nature Machine Intelligence}\ }\textbf {\bibinfo
  {volume} {7}},\ \bibinfo {pages} {836} (\bibinfo {year} {2025})}\BibitemShut
  {NoStop}%
\bibitem [{\citenamefont {Sun}\ \emph {et~al.}(2016)\citenamefont {Sun},
  \citenamefont {Dacek}, \citenamefont {Ong}, \citenamefont {Hautier},
  \citenamefont {Jain}, \citenamefont {Richards}, \citenamefont {Gamst},
  \citenamefont {Persson},\ and\ \citenamefont {Ceder}}]{sun2016thermodynamic}%
  \BibitemOpen
  \bibfield  {author} {\bibinfo {author} {\bibfnamefont {W.}~\bibnamefont
  {Sun}}, \bibinfo {author} {\bibfnamefont {S.~T.}\ \bibnamefont {Dacek}},
  \bibinfo {author} {\bibfnamefont {S.~P.}\ \bibnamefont {Ong}}, \bibinfo
  {author} {\bibfnamefont {G.}~\bibnamefont {Hautier}}, \bibinfo {author}
  {\bibfnamefont {A.}~\bibnamefont {Jain}}, \bibinfo {author} {\bibfnamefont
  {W.~D.}\ \bibnamefont {Richards}}, \bibinfo {author} {\bibfnamefont {A.~C.}\
  \bibnamefont {Gamst}}, \bibinfo {author} {\bibfnamefont {K.~A.}\ \bibnamefont
  {Persson}},\ and\ \bibinfo {author} {\bibfnamefont {G.}~\bibnamefont
  {Ceder}},\ }\bibfield  {title} {\bibinfo {title} {The thermodynamic scale of
  inorganic crystalline metastability},\ }\href
  {https://www.science.org/doi/10.1126/sciadv.1600225} {\bibfield  {journal}
  {\bibinfo  {journal} {Science advances}\ }\textbf {\bibinfo {volume} {2}},\
  \bibinfo {pages} {e1600225} (\bibinfo {year} {2016})}\BibitemShut {NoStop}%
\bibitem [{\citenamefont {Silver}\ and\ \citenamefont
  {Sutton}(2025)}]{silver2025welcome}%
  \BibitemOpen
  \bibfield  {author} {\bibinfo {author} {\bibfnamefont {D.}~\bibnamefont
  {Silver}}\ and\ \bibinfo {author} {\bibfnamefont {R.~S.}\ \bibnamefont
  {Sutton}},\ }\bibfield  {title} {\bibinfo {title} {Welcome to the era of
  experience},\ }\href
  {http://incompleteideas.net/papers/TheEraOfExperience.pdf} {\bibfield
  {journal} {\bibinfo  {journal} {Google AI}\ }\textbf {\bibinfo {volume} {1}}
  (\bibinfo {year} {2025})}\BibitemShut {NoStop}%
\bibitem [{\citenamefont {Cao}\ and\ \citenamefont
  {Wang}(2025)}]{crystalformer-rl}%
  \BibitemOpen
  \bibfield  {author} {\bibinfo {author} {\bibfnamefont {Z.}~\bibnamefont
  {Cao}}\ and\ \bibinfo {author} {\bibfnamefont {L.}~\bibnamefont {Wang}},\
  }\href@noop {} {\bibinfo {title} {Crystalformer-rl: Reinforcement fine-tuning
  for materials design}} (\bibinfo {year} {2025}),\ \Eprint
  {https://arxiv.org/abs/2504.02367} {arXiv:2504.02367 [cond-mat.mtrl-sci]}
  \BibitemShut {NoStop}%
\bibitem [{\citenamefont {Shenfeld}\ \emph {et~al.}(2025)\citenamefont
  {Shenfeld}, \citenamefont {Pari},\ and\ \citenamefont
  {Agrawal}}]{shenfeld2025rlsrazoronlinereinforcement}%
  \BibitemOpen
  \bibfield  {author} {\bibinfo {author} {\bibfnamefont {I.}~\bibnamefont
  {Shenfeld}}, \bibinfo {author} {\bibfnamefont {J.}~\bibnamefont {Pari}},\
  and\ \bibinfo {author} {\bibfnamefont {P.}~\bibnamefont {Agrawal}},\ }\href
  {https://arxiv.org/abs/2509.04259} {\bibinfo {title} {Rl's razor: Why online
  reinforcement learning forgets less}} (\bibinfo {year} {2025}),\ \Eprint
  {https://arxiv.org/abs/2509.04259} {arXiv:2509.04259 [cs.LG]} \BibitemShut
  {NoStop}%
\bibitem [{\citenamefont {Batsheva}\ \emph {et~al.}(2023)\citenamefont
  {Batsheva}, \citenamefont {Chertkov}, \citenamefont {Ryzhakov},\ and\
  \citenamefont {Oseledets}}]{PROTES}%
  \BibitemOpen
  \bibfield  {author} {\bibinfo {author} {\bibfnamefont {A.}~\bibnamefont
  {Batsheva}}, \bibinfo {author} {\bibfnamefont {A.}~\bibnamefont {Chertkov}},
  \bibinfo {author} {\bibfnamefont {G.}~\bibnamefont {Ryzhakov}},\ and\
  \bibinfo {author} {\bibfnamefont {I.}~\bibnamefont {Oseledets}},\ }\bibfield
  {title} {\bibinfo {title} {Protes: Probabilistic optimization with tensor
  sampling},\ }in\ \href
  {https://proceedings.neurips.cc/paper_files/paper/2023/file/028957869e560af14243ac37663a471e-Paper-Conference.pdf}
  {\emph {\bibinfo {booktitle} {Advances in Neural Information Processing
  Systems}}},\ Vol.~\bibinfo {volume} {36},\ \bibinfo {editor} {edited by\
  \bibinfo {editor} {\bibfnamefont {A.}~\bibnamefont {Oh}}, \bibinfo {editor}
  {\bibfnamefont {T.}~\bibnamefont {Naumann}}, \bibinfo {editor} {\bibfnamefont
  {A.}~\bibnamefont {Globerson}}, \bibinfo {editor} {\bibfnamefont
  {K.}~\bibnamefont {Saenko}}, \bibinfo {editor} {\bibfnamefont
  {M.}~\bibnamefont {Hardt}},\ and\ \bibinfo {editor} {\bibfnamefont
  {S.}~\bibnamefont {Levine}}}\ (\bibinfo  {publisher} {Curran Associates,
  Inc.},\ \bibinfo {year} {2023})\ pp.\ \bibinfo {pages} {808--823}\BibitemShut
  {NoStop}%
\bibitem [{\citenamefont {Sozykin}\ \emph {et~al.}(2025)\citenamefont
  {Sozykin}, \citenamefont {Chertkov}, \citenamefont {Phan}, \citenamefont
  {Oseledets},\ and\ \citenamefont
  {Ryzhakov}}]{sozykin2025highdimensionaloptimizationlowrank}%
  \BibitemOpen
  \bibfield  {author} {\bibinfo {author} {\bibfnamefont {K.}~\bibnamefont
  {Sozykin}}, \bibinfo {author} {\bibfnamefont {A.}~\bibnamefont {Chertkov}},
  \bibinfo {author} {\bibfnamefont {A.-H.}\ \bibnamefont {Phan}}, \bibinfo
  {author} {\bibfnamefont {I.}~\bibnamefont {Oseledets}},\ and\ \bibinfo
  {author} {\bibfnamefont {G.}~\bibnamefont {Ryzhakov}},\ }\href
  {https://arxiv.org/abs/2505.12383} {\bibinfo {title} {High-dimensional
  optimization with low rank tensor sampling and local search}} (\bibinfo
  {year} {2025}),\ \Eprint {https://arxiv.org/abs/2505.12383} {arXiv:2505.12383
  [math.OC]} \BibitemShut {NoStop}%
\bibitem [{\citenamefont {Ouyang}\ \emph {et~al.}(2022)\citenamefont {Ouyang},
  \citenamefont {Wu}, \citenamefont {Jiang}, \citenamefont {Almeida},
  \citenamefont {Wainwright}, \citenamefont {Mishkin}, \citenamefont {Zhang},
  \citenamefont {Agarwal}, \citenamefont {Slama}, \citenamefont {Ray},
  \citenamefont {Schulman}, \citenamefont {Hilton}, \citenamefont {Kelton},
  \citenamefont {Miller}, \citenamefont {Simens}, \citenamefont {Askell},
  \citenamefont {Welinder}, \citenamefont {Christiano}, \citenamefont {Leike},\
  and\ \citenamefont {Lowe}}]{Ouyang2022InstructGPT}%
  \BibitemOpen
  \bibfield  {author} {\bibinfo {author} {\bibfnamefont {L.}~\bibnamefont
  {Ouyang}}, \bibinfo {author} {\bibfnamefont {J.}~\bibnamefont {Wu}}, \bibinfo
  {author} {\bibfnamefont {X.}~\bibnamefont {Jiang}}, \bibinfo {author}
  {\bibfnamefont {D.~F.}\ \bibnamefont {Almeida}}, \bibinfo {author}
  {\bibfnamefont {C.~L.}\ \bibnamefont {Wainwright}}, \bibinfo {author}
  {\bibfnamefont {P.}~\bibnamefont {Mishkin}}, \bibinfo {author} {\bibfnamefont
  {C.}~\bibnamefont {Zhang}}, \bibinfo {author} {\bibfnamefont
  {S.}~\bibnamefont {Agarwal}}, \bibinfo {author} {\bibfnamefont
  {K.}~\bibnamefont {Slama}}, \bibinfo {author} {\bibfnamefont
  {A.}~\bibnamefont {Ray}}, \bibinfo {author} {\bibfnamefont {J.}~\bibnamefont
  {Schulman}}, \bibinfo {author} {\bibfnamefont {J.}~\bibnamefont {Hilton}},
  \bibinfo {author} {\bibfnamefont {F.}~\bibnamefont {Kelton}}, \bibinfo
  {author} {\bibfnamefont {L.~E.}\ \bibnamefont {Miller}}, \bibinfo {author}
  {\bibfnamefont {M.}~\bibnamefont {Simens}}, \bibinfo {author} {\bibfnamefont
  {A.}~\bibnamefont {Askell}}, \bibinfo {author} {\bibfnamefont
  {P.}~\bibnamefont {Welinder}}, \bibinfo {author} {\bibfnamefont
  {P.}~\bibnamefont {Christiano}}, \bibinfo {author} {\bibfnamefont
  {J.}~\bibnamefont {Leike}},\ and\ \bibinfo {author} {\bibfnamefont {R.~J.}\
  \bibnamefont {Lowe}},\ }\bibfield  {title} {\bibinfo {title} {Training
  language models to follow instructions with human feedback},\ }in\ \href
  {https://proceedings.neurips.cc/paper/2022/file/b1efde53be364a73914f58805a001731-Paper.pdf}
  {\emph {\bibinfo {booktitle} {Advances in Neural Information Processing
  Systems 35 (NeurIPS 2022) Workshops}}}\ (\bibinfo {year} {2022})\ \bibinfo
  {note} {originally submitted as arXiv:2203.02155}\BibitemShut {NoStop}%
\bibitem [{\citenamefont {Schulman}\ \emph {et~al.}(2017)\citenamefont
  {Schulman}, \citenamefont {Wolski}, \citenamefont {Dhariwal}, \citenamefont
  {Radford},\ and\ \citenamefont {Klimov}}]{schulman2017ppo}%
  \BibitemOpen
  \bibfield  {author} {\bibinfo {author} {\bibfnamefont {J.}~\bibnamefont
  {Schulman}}, \bibinfo {author} {\bibfnamefont {F.}~\bibnamefont {Wolski}},
  \bibinfo {author} {\bibfnamefont {P.}~\bibnamefont {Dhariwal}}, \bibinfo
  {author} {\bibfnamefont {A.}~\bibnamefont {Radford}},\ and\ \bibinfo {author}
  {\bibfnamefont {O.}~\bibnamefont {Klimov}},\ }\bibfield  {title} {\bibinfo
  {title} {Proximal policy optimization algorithms},\ }\href
  {https://arxiv.org/abs/1707.06347} {\bibfield  {journal} {\bibinfo  {journal}
  {CoRR}\ }\textbf {\bibinfo {volume} {abs/1707.06347}} (\bibinfo {year}
  {2017})}\BibitemShut {NoStop}%
\bibitem [{\citenamefont {Mirdita}\ \emph {et~al.}(2022)\citenamefont
  {Mirdita}, \citenamefont {Sch{\"u}tze}, \citenamefont {Moriwaki},
  \citenamefont {Heo}, \citenamefont {Ovchinnikov},\ and\ \citenamefont
  {Steinegger}}]{Mirdita2022}%
  \BibitemOpen
  \bibfield  {author} {\bibinfo {author} {\bibfnamefont {M.}~\bibnamefont
  {Mirdita}}, \bibinfo {author} {\bibfnamefont {K.}~\bibnamefont
  {Sch{\"u}tze}}, \bibinfo {author} {\bibfnamefont {Y.}~\bibnamefont
  {Moriwaki}}, \bibinfo {author} {\bibfnamefont {L.}~\bibnamefont {Heo}},
  \bibinfo {author} {\bibfnamefont {S.}~\bibnamefont {Ovchinnikov}},\ and\
  \bibinfo {author} {\bibfnamefont {M.}~\bibnamefont {Steinegger}},\ }\bibfield
   {title} {\bibinfo {title} {Colabfold: making protein folding accessible to
  all},\ }\href {https://doi.org/10.1038/s41592-022-01488-1} {\bibfield
  {journal} {\bibinfo  {journal} {Nature Methods}\ }\textbf {\bibinfo {volume}
  {19}},\ \bibinfo {pages} {679} (\bibinfo {year} {2022})}\BibitemShut
  {NoStop}%
\bibitem [{\citenamefont {Liu}\ \emph {et~al.}(2024)\citenamefont {Liu},
  \citenamefont {Tamaki}, \citenamefont {Yokoyama}, \citenamefont {Wakasugi},
  \citenamefont {Yotsuhashi}, \citenamefont {Kusaba}, \citenamefont {Oganov},\
  and\ \citenamefont {Yoshida}}]{shotguncsp}%
  \BibitemOpen
  \bibfield  {author} {\bibinfo {author} {\bibfnamefont {C.}~\bibnamefont
  {Liu}}, \bibinfo {author} {\bibfnamefont {H.}~\bibnamefont {Tamaki}},
  \bibinfo {author} {\bibfnamefont {T.}~\bibnamefont {Yokoyama}}, \bibinfo
  {author} {\bibfnamefont {K.}~\bibnamefont {Wakasugi}}, \bibinfo {author}
  {\bibfnamefont {S.}~\bibnamefont {Yotsuhashi}}, \bibinfo {author}
  {\bibfnamefont {M.}~\bibnamefont {Kusaba}}, \bibinfo {author} {\bibfnamefont
  {A.~R.}\ \bibnamefont {Oganov}},\ and\ \bibinfo {author} {\bibfnamefont
  {R.}~\bibnamefont {Yoshida}},\ }\bibfield  {title} {\bibinfo {title} {Shotgun
  crystal structure prediction using machine-learned formation energies},\
  }\href {https://www.nature.com/articles/s41524-024-01471-8} {\bibfield
  {journal} {\bibinfo  {journal} {npj Computational Materials}\ }\textbf
  {\bibinfo {volume} {10}},\ \bibinfo {pages} {298} (\bibinfo {year}
  {2024})}\BibitemShut {NoStop}%
\bibitem [{Bat()}]{BatchRelaxer}%
  \BibitemOpen
  \href@noop {} {}\bibinfo {note}
  {\url{https://github.com/zdcao121/BatchRelaxer}}\BibitemShut {NoStop}%
\bibitem [{\citenamefont {Wang}\ \emph
  {et~al.}(2025{\natexlab{b}})\citenamefont {Wang}, \citenamefont {Wang},
  \citenamefont {Wang}, \citenamefont {Wu}, \citenamefont {Lv},\ and\
  \citenamefont {Wang}}]{opencsp}%
  \BibitemOpen
  \bibfield  {author} {\bibinfo {author} {\bibfnamefont {Y.}~\bibnamefont
  {Wang}}, \bibinfo {author} {\bibfnamefont {X.}~\bibnamefont {Wang}}, \bibinfo
  {author} {\bibfnamefont {Z.}~\bibnamefont {Wang}}, \bibinfo {author}
  {\bibfnamefont {J.}~\bibnamefont {Wu}}, \bibinfo {author} {\bibfnamefont
  {J.}~\bibnamefont {Lv}},\ and\ \bibinfo {author} {\bibfnamefont
  {H.}~\bibnamefont {Wang}},\ }\href {https://arxiv.org/abs/2509.10293}
  {\bibinfo {title} {Opencsp: A deep learning framework for crystal structure
  prediction from ambient to high pressure}} (\bibinfo {year}
  {2025}{\natexlab{b}}),\ \Eprint {https://arxiv.org/abs/2509.10293}
  {arXiv:2509.10293 [cond-mat.mtrl-sci]} \BibitemShut {NoStop}%
\bibitem [{\citenamefont {Kruglov}\ \emph {et~al.}(2023)\citenamefont
  {Kruglov}, \citenamefont {Yanilkin}, \citenamefont {Propad}, \citenamefont
  {Mazitov}, \citenamefont {Rachitskii},\ and\ \citenamefont
  {Oganov}}]{kruglov2023crystal}%
  \BibitemOpen
  \bibfield  {author} {\bibinfo {author} {\bibfnamefont {I.~A.}\ \bibnamefont
  {Kruglov}}, \bibinfo {author} {\bibfnamefont {A.~V.}\ \bibnamefont
  {Yanilkin}}, \bibinfo {author} {\bibfnamefont {Y.}~\bibnamefont {Propad}},
  \bibinfo {author} {\bibfnamefont {A.~B.}\ \bibnamefont {Mazitov}}, \bibinfo
  {author} {\bibfnamefont {P.}~\bibnamefont {Rachitskii}},\ and\ \bibinfo
  {author} {\bibfnamefont {A.~R.}\ \bibnamefont {Oganov}},\ }\bibfield  {title}
  {\bibinfo {title} {Crystal structure prediction at finite temperatures},\
  }\href {https://www.nature.com/articles/s41524-023-01120-6} {\bibfield
  {journal} {\bibinfo  {journal} {npj Computational Materials}\ }\textbf
  {\bibinfo {volume} {9}},\ \bibinfo {pages} {197} (\bibinfo {year}
  {2023})}\BibitemShut {NoStop}%
\bibitem [{\citenamefont {O’keeffe}\ \emph {et~al.}(2008)\citenamefont
  {O’keeffe}, \citenamefont {Peskov}, \citenamefont {Ramsden},\ and\
  \citenamefont {Yaghi}}]{o2008reticular}%
  \BibitemOpen
  \bibfield  {author} {\bibinfo {author} {\bibfnamefont {M.}~\bibnamefont
  {O’keeffe}}, \bibinfo {author} {\bibfnamefont {M.~A.}\ \bibnamefont
  {Peskov}}, \bibinfo {author} {\bibfnamefont {S.~J.}\ \bibnamefont
  {Ramsden}},\ and\ \bibinfo {author} {\bibfnamefont {O.~M.}\ \bibnamefont
  {Yaghi}},\ }\bibfield  {title} {\bibinfo {title} {The reticular chemistry
  structure resource (rcsr) database of, and symbols for, crystal nets},\
  }\href {https://pubs.acs.org/doi/10.1021/ar800124u} {\bibfield  {journal}
  {\bibinfo  {journal} {Accounts of chemical research}\ }\textbf {\bibinfo
  {volume} {41}},\ \bibinfo {pages} {1782} (\bibinfo {year}
  {2008})}\BibitemShut {NoStop}%
\bibitem [{\citenamefont {Leeman}\ \emph {et~al.}(2024)\citenamefont {Leeman},
  \citenamefont {Liu}, \citenamefont {Stiles}, \citenamefont {Lee},
  \citenamefont {Bhatt}, \citenamefont {Schoop},\ and\ \citenamefont
  {Palgrave}}]{PRXEnergy.3.011002}%
  \BibitemOpen
  \bibfield  {author} {\bibinfo {author} {\bibfnamefont {J.}~\bibnamefont
  {Leeman}}, \bibinfo {author} {\bibfnamefont {Y.}~\bibnamefont {Liu}},
  \bibinfo {author} {\bibfnamefont {J.}~\bibnamefont {Stiles}}, \bibinfo
  {author} {\bibfnamefont {S.~B.}\ \bibnamefont {Lee}}, \bibinfo {author}
  {\bibfnamefont {P.}~\bibnamefont {Bhatt}}, \bibinfo {author} {\bibfnamefont
  {L.~M.}\ \bibnamefont {Schoop}},\ and\ \bibinfo {author} {\bibfnamefont
  {R.~G.}\ \bibnamefont {Palgrave}},\ }\bibfield  {title} {\bibinfo {title}
  {Challenges in high-throughput inorganic materials prediction and autonomous
  synthesis},\ }\href {https://doi.org/10.1103/PRXEnergy.3.011002} {\bibfield
  {journal} {\bibinfo  {journal} {PRX Energy}\ }\textbf {\bibinfo {volume}
  {3}},\ \bibinfo {pages} {011002} (\bibinfo {year} {2024})}\BibitemShut
  {NoStop}%
\bibitem [{\citenamefont {Cheetham}\ and\ \citenamefont
  {Seshadri}(2024)}]{cheetham2024artificial}%
  \BibitemOpen
  \bibfield  {author} {\bibinfo {author} {\bibfnamefont {A.~K.}\ \bibnamefont
  {Cheetham}}\ and\ \bibinfo {author} {\bibfnamefont {R.}~\bibnamefont
  {Seshadri}},\ }\bibfield  {title} {\bibinfo {title} {Artificial intelligence
  driving materials discovery? perspective on the article: Scaling deep
  learning for materials discovery},\ }\href
  {https://pubs.acs.org/doi/10.1021/acs.chemmater.4c00643} {\bibfield
  {journal} {\bibinfo  {journal} {Chemistry of Materials}\ }\textbf {\bibinfo
  {volume} {36}},\ \bibinfo {pages} {3490} (\bibinfo {year}
  {2024})}\BibitemShut {NoStop}%
\bibitem [{\citenamefont {Ho}\ and\ \citenamefont
  {Salimans}(2022)}]{ho2022classifier}%
  \BibitemOpen
  \bibfield  {author} {\bibinfo {author} {\bibfnamefont {J.}~\bibnamefont
  {Ho}}\ and\ \bibinfo {author} {\bibfnamefont {T.}~\bibnamefont {Salimans}},\
  }\bibfield  {title} {\bibinfo {title} {Classifier-free diffusion guidance},\
  }\href {https://arxiv.org/abs/2207.12598} {\  (\bibinfo {year} {2022})},\
  \Eprint {https://arxiv.org/abs/2207.12598} {arXiv:2207.12598 [cs.LG]}
  \BibitemShut {NoStop}%
\bibitem [{\citenamefont {Kahneman}(2011)}]{kahneman2011thinking}%
  \BibitemOpen
  \bibfield  {author} {\bibinfo {author} {\bibfnamefont {D.}~\bibnamefont
  {Kahneman}},\ }\href
  {https://ia600603.us.archive.org/10/items/DanielKahnemanThinkingFastAndSlow/Daniel%20Kahneman-Thinking%2C%20Fast%20and%20Slow%20%20.pdf}
  {\emph {\bibinfo {title} {Thinking, Fast and Slow}}}\ (\bibinfo  {publisher}
  {Farrar, Straus and Giroux},\ \bibinfo {address} {New York},\ \bibinfo {year}
  {2011})\BibitemShut {NoStop}%
\end{thebibliography}%

\end{document}